\documentclass[10pt,journal,compsoc]{IEEEtran}
\ifCLASSOPTIONcompsoc
  \usepackage[nocompress]{cite}
\else
  \usepackage{cite}
\fi
\ifCLASSINFOpdf
\else
\fi
\usepackage{subfigure}
\usepackage{wrapfig}
\usepackage{amsmath}
\usepackage{amssymb}
\DeclareMathOperator*{\argmin}{arg\,min}
\DeclareMathOperator*{\argmax}{arg\,max}
\usepackage{multirow}
\usepackage{multicol}
\usepackage{graphicx}
\usepackage{caption}
\usepackage{makecell}
\usepackage{xcolor}
\usepackage[normalem]{ulem}

\usepackage[ruled]{algorithm2e} 

\SetAlFnt{\small}
\SetAlCapFnt{\small}
\SetAlCapNameFnt{\small}
\SetAlCapHSkip{0pt}

\usepackage{color}
\definecolor{blue}{rgb}{0,0,0}
\definecolor{green}{rgb}{0,0,0}
\definecolor{red}{rgb}{0.7,0,0}
\def\blue{\color{blue}}

\begin{document}

%
\title{Placement Retargeting of Virtual Avatars to Dissimilar Indoor Environments}
%
%
%
%

\author{Leonard Yoon,~\IEEEmembership{Student Member,~IEEE,}
        Dongseok Yang,~\IEEEmembership{Student Member,~IEEE,}
        Jaehyun Kim,
        Choongho Chung,~\IEEEmembership{Student Member,~IEEE,}
        Sung-Hee Lee,~\IEEEmembership{Member,~IEEE}
\IEEEcompsocitemizethanks{\IEEEcompsocthanksitem Leonard Yoon, Dongseok Yang, Jaehyun Kim, Choongho Chung and Sung-Hee Lee are with Korea Advanced Institute of Science and Technology (KAIST).\protect\\
E-mail: \{lyoon, dsyang, chrisjkim, thegenuine, sunghee.lee\}@kaist.ac.kr
}
\thanks{Manuscript received March 9, 2020; revised August 19, 2020.}}

%
%

\markboth{TRANSACTIONS ON VISUALIZATION AND COMPUTER GRAPHICS, AUGUST 2020}%
{Shell \MakeLowercase{\textit{et al.}}: Bare Demo of IEEEtran.cls for Computer Society Journals}
%



 
\IEEEtitleabstractindextext{%
\setcounter{figure}{0} 
\begin{center}
    \centering
    \includegraphics[width=1.5\columnwidth]{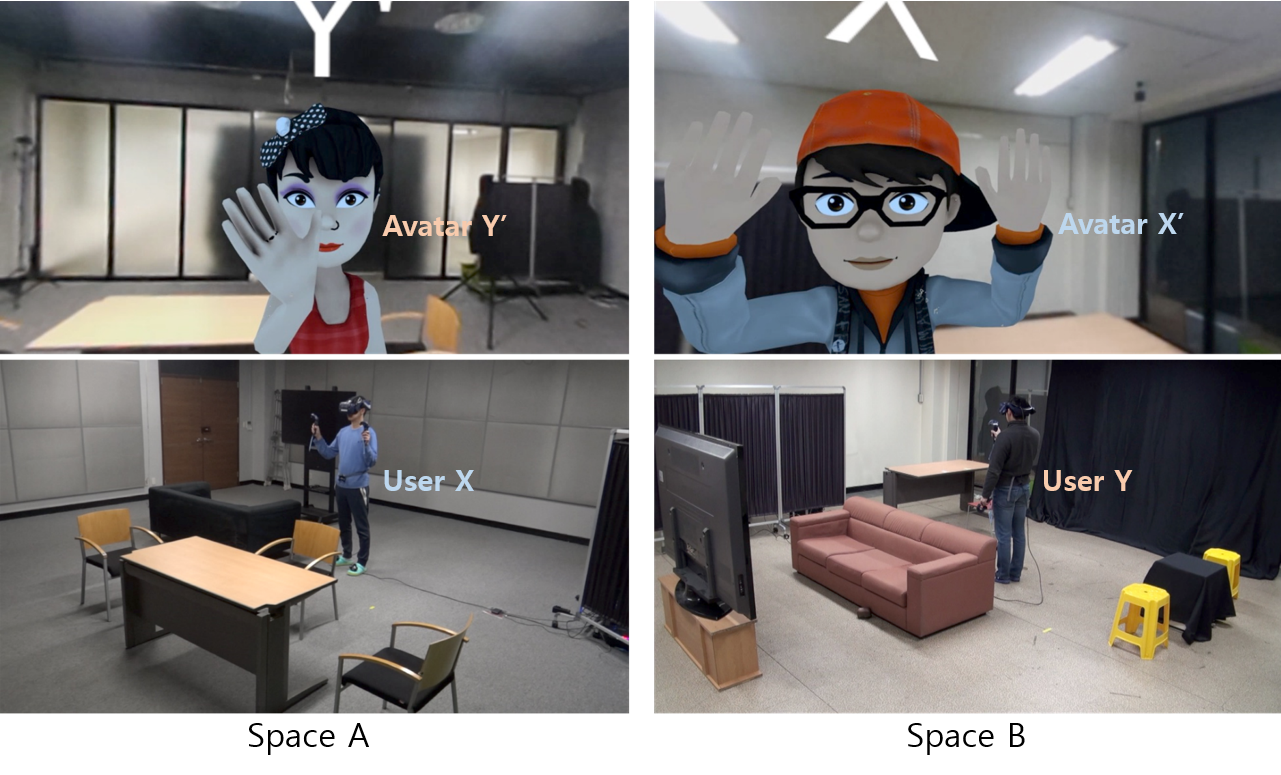}
    \captionof{figure}{{\blue AR telepresence spaces seen from the perspective views (bottom) of the User $X$ in Space A (left) and User $Y$ in Space B (right), and the egocentric views of the users seeing the avatar of the other party (top). The avatars are placed by our method to preserve the semantics of the placement the other party.}}
    \label{fig:teaser} 
\end{center}

\begin{abstract}
Rapidly developing technologies are realizing a 3D telepresence, in which geographically separated users can interact with each other through their virtual avatars. In this paper, we present novel methods to determine the avatar's position in an indoor space to preserve the semantics of the user's position in a dissimilar indoor space with different space configurations and furniture layouts. To this end, we first perform a user survey on the preferred avatar placements for various indoor configurations and user placements, and identify a set of related attributes, including interpersonal relation, visual attention, pose, and spatial characteristics, and quantify these attributes with a set of features. By using the obtained dataset and identified features, we train a neural network that predicts the similarity between two placements. Next, we develop an avatar placement method that preserves the semantics of the placement of the remote user in a different space as much as possible. We show the effectiveness of our methods by implementing a prototype  AR-based telepresence system and user evaluations.
\end{abstract}

}

\maketitle

\IEEEdisplaynontitleabstractindextext

%
\IEEEpeerreviewmaketitle

\IEEEraisesectionheading{\section{Introduction}\label{sec:introduction}}

Rapidly advancing technologies are gradually realizing immersive 3D telepresence experience, in which a user can interact with a remote user through his/her virtual avatar, an image of the remote user overlaid on the local physical space, thereby significantly improving the sense of co-presence \cite{maimone2013general}. For two-way communication, the local user is also represented by an avatar to be seen by the remote user.

In this avatar-mediated telepresence, virtual avatar needs to be placed and animated to deliver the semantics of the remote user's movement to the local user.
If the configurations of the two remote spaces are identical, which allows for identifying a rigid transformation that maps between the two spaces, it can be achieved by capturing the user's placement and motion and applying them to his/her virtual avatar at the corresponding remote placement \cite{holoporation}.

However, due to the sheer variety of human living spaces, two users are more likely in rooms with different shape and furniture arrangement, where simple mapping between the placement of two spaces does not exist. 
In this case, one needs a non-trivial method that determines the placement and movement of avatar according to the configuration of the local environment so that the local user can properly recognize what the remote user is doing and interact with him/her through avatar.

Several methods have been developed to tackle this problem, e.g., finding an affine transformation that maximizes shared free spaces \cite{Lehment2014} or matching sittable regions between two spaces \cite{Pejsa:2016, Jo:2015} for teleconference. 
However, these methods allow only sub-regions within whole space to be used for telepresence, and do not provide mapping from an arbitrary placement of one space to the other. In this paper, we aim to develop a method that allows the users to fully utilize the two indoor spaces for the purpose of room-scale telepresence.

Placing and animating telepresence avatar to deliver the remote user's motion semantics must consider various factors of human motion in real life, including the user's motion, attention, and the relationship to nearby people or objects. On the higher level, other important aspects should be considered as well, such as the naturalness of avatar animation, guarantees of availability and reachability of interaction resources, and user immersion. Because of the difficulty of recognizing some latent attributes and compromising among a number of factors, the avatar placement and animation problem to best preserve the semantics of the original motion is a challenging task, with its difficulty increasing with the dissimilarity between the two environments. Among a number of issues discussed above, this paper deals with the problem of determining the placement of an avatar in a room-scale indoor environment.

\begin{figure}
\includegraphics[width=1\columnwidth]{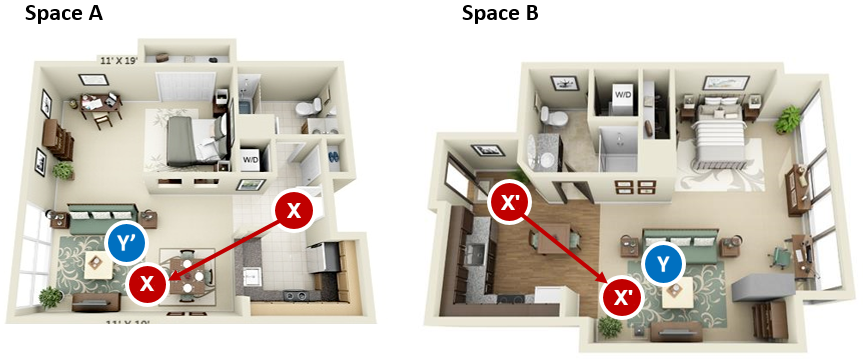}
  \caption{Our target scenario for 3D telepresence. Person $X$ in Space A and Person $Y$ in Space B are experiencing co-presence through avatars $X'$ and $Y'$ representing remote users. When Person $X$ moves to a new placement, our method determines the optimal placement of Avatar $X'$ to best preserves the semantics of the placement of users.}
  \label{fig:scene}  
\end{figure}

\subsection{Problem Definition and Our Approach}
\label{sec:problem_definition}

The telepresence scenario of our research is illustrated in Figure \ref{fig:scene} showing two remote rooms that have different layouts and furniture arrangements. By teleporting a Person $X$ in Space A to Space B as an Avatar $X'$, a Person $Y$ in Space B with an HMD can see $X'$. Likewise, $X$ can see $Y$'s avatar $Y'$ in Space A; this allows each user to feel that the remote user is visiting his/her place and interact with him/her as if they were in the same space. Furthermore, we assume the ``everyday telepresence'' \cite{Rae:2015:ETE:2702613.2702639,Schlager:2015}, in which two people live a daily life in the telepresence environment: they interact with each other at times and also may act independently. Now, suppose that $X$ has moved to a certain position within Space A. Then where should $X'$ in Space B be located to best represent $X$?

This question is related to understanding the semantics of $X$ being at $p_X$ in Space A, where $p_{X}$ denotes the placement (position and orientation) of $X$, and for this a number of different attributes need to be considered:
\begin{itemize}
\item {\bf Interaction:} $X$ may be at $p_X$ in order to interact with a person or object near $p_X$. In this case, $X'$ needs to be placed where the same interaction is possible.
\item {\bf Pose:} $X$ can take a particular pose afforded by $p_X$ (e.g., sitting on a chair), and it would be ideal to place $X'$ where the same pose can be accommodated.
\item {\bf Space:} An indoor space is divided into different functional spaces, such as dining and studying areas. $X'$ may need to be placed in the same functional space as $X$.
\end{itemize}
In this paper, we develop a set of features that quantify these attributes.

After defining such features, we develop a method to find an appropriate corresponding placement of $X'$.
This can be formulated as an optimization problem defined as: 
\begin{equation}
\argmax_{p_{X'}} Sim( x( p_X | p_{Y'}, A), x( p_{X'} | p_Y, B) ),
\end{equation}
where $x(p_{X'} | p_Y, B)$ denotes the feature vector of $p_{X'}$ representing its semantics given $p_Y$ and Space B. Then finding an appropriate similarity function $Sim(\cdot,\cdot)$, which measures how close the two features are from the viewpoint of the telepresence, is a key for the avatar placement.

If the shape and furniture arrangement of two distant spaces around $Y$ and $Y'$ are similar, all the features will point to similar locations for the placement of $X'$. Otherwise, each feature may give higher preference to different locations. For instance, when two persons communicate, interaction-related features may dominate other features while pose and space-related features will be more important when users act independently. Consequently, learning the similarity measure is related to estimating the relative importance among the features, and the relative importance varies with situations. Therefore, the similarity measure will be a non-linear function of the features. 

As the similarity measure is in the realm of human perception, we take the approach to learn the measure with a dataset obtained by a user survey on how users would locate their avatars in various situations. In particular, we take a ranking approach to solve the optimization problem and trained a neural network to estimate the similarity of two placements in different spaces using a triplet loss architecture. The trained neural network surpasses a baseline linear model by a large margin in terms of test accuracy.

To investigate the effectiveness of our method, we construct a prototype telepresence system and conduct a user study. Figure \ref{fig:teaser} shows a screenshot of the developed system.
The user study shows that several placement qualities of our method are comparable to those of manual placement when evaluating the avatar of the other party in our test AR telepresence environments.

{\bf Contributions.} 
We develop a novel framework that determines the placement of a virtual avatar in a general indoor space to preserve various aspects of the meaning of the user's placement in a dissimilar remote space. To this end, we identified a set of features to be considered for the placement and developed a neural network-based similarity predictor by using a training set obtained from a user survey. We developed a prototype AR-based telepresence system and show the effectiveness of our methods.

The remainder of the paper proceeds as follows. After reviewing related work in Section \ref{sec:related_work}, we present our method for training a similarity predictor between two placements for generating the placement of the virtual avatar in Section \ref{sec:method}. We evaluate our method with respect to the prediction accuracy and a user study by using the developed prototype telepresence system in Section \ref{sec:result}. Section \ref{sec:discussion} concludes the paper with discussion of the limitations of our research and future research directions.
\section{Related Work}
\label{sec:related_work}
In this section, we review related studies on the placement of the telepresence avatars, motion retargeting, and understanding the  relationships between a character and nearby entities.
\subsection{Placement of Telepresence Avatar}
Recent technical advances have facilitated bidirectional immersive telepresence, which merges two remote spaces into one. 

Maimone and Fuchs \cite{maimone2013general} proposed an HMD-based telepresence system that overlays a remote user's image on a local space. Microsoft Holoportation transmits 3D data of remote people and objects to a local space \cite{holoporation}. These studies focused on capturing the remote person and reconstructing him/her in a designated local space, but they did not consider the problems arising from the differences between two spaces, which is the main problem of our research.

This problem was partly addressed by Pejsa et al. \cite{Pejsa:2016} who developed a system that projects a remote user's image at a certain local position for which correspondence is manually specified (e.g., sofa to sofa). Jo et al. \cite{Jo:2015} proposed a method to select a suitable furniture object that an avatar should be located near by constructing the correspondence between objects in two spaces and adjusted the body pose to fit the different shapes of the object. Lehment et al. \cite{Lehment2014} proposed a  method that rigidly aligns two spaces to maximally overlap free spaces and work surfaces so that a remote person in a free space could be transmitted to a free space. However, the rigid alignment approach is valid only if the two spaces are similar. By contrast, we consider general floor layouts that a rigid alignment method cannot handle. More importantly, we find the corresponding positions by considering many important factors, such as the attention and pose of a person, the spatial relation between people, and spatial characteristics of spaces.

In recent years, there have been work on mixed reality (MR) system that adaptively positioned a resized avatar within the field of view of AR HMD for remote collaboration \cite{piumsomboon2018mini}, and proposed a system realizing transitions between VR and AR to provide video, audio and spatial capture for teaching physical tasks \cite{thoravi2019loki}. They alleviated the problem arose from hardware and extended ability of MR system with integration of existing technology. While these studies focus on either the interaction only in shared space for collaboration or the system that requires additional user input for teaching task, we deal with general use of everyday telepresence for people naturally living in different spaces.

\subsection{Retargeting Human Movement}
Our problem shares the same goal with the motion retargeting problem in that an existing human motion is modified to match different situations. 
Early methods mostly solved the problem of modifying a given motion for new characters with different body dimensions or varied environments \cite{Gleicher:1998:RMN, choi2000online}. When it comes to the interaction motions between two people or between a person and an object, the retargeted motion can be meaningful only if the interaction semantics between the two is preserved. For this, many approaches have defined the spatial relationships between the interacting entities \cite{Ho:2010:SRP,Al-Asqhar:2013:RDI,Wang:2013:HPE,Kim2016,Jin:2018}.

Retargeting of a person's placement to other environments in an everyday scenario should also preserve the semantics of the person's location. Compared with previous motion retargeting studies, this problem must consider significantly more factors involved in human motion as mentioned above. This paper takes these factors into consideration to obtain the avatar placement method.

\begin{figure*}[t]
\centering
\includegraphics[width=1.6\columnwidth]
{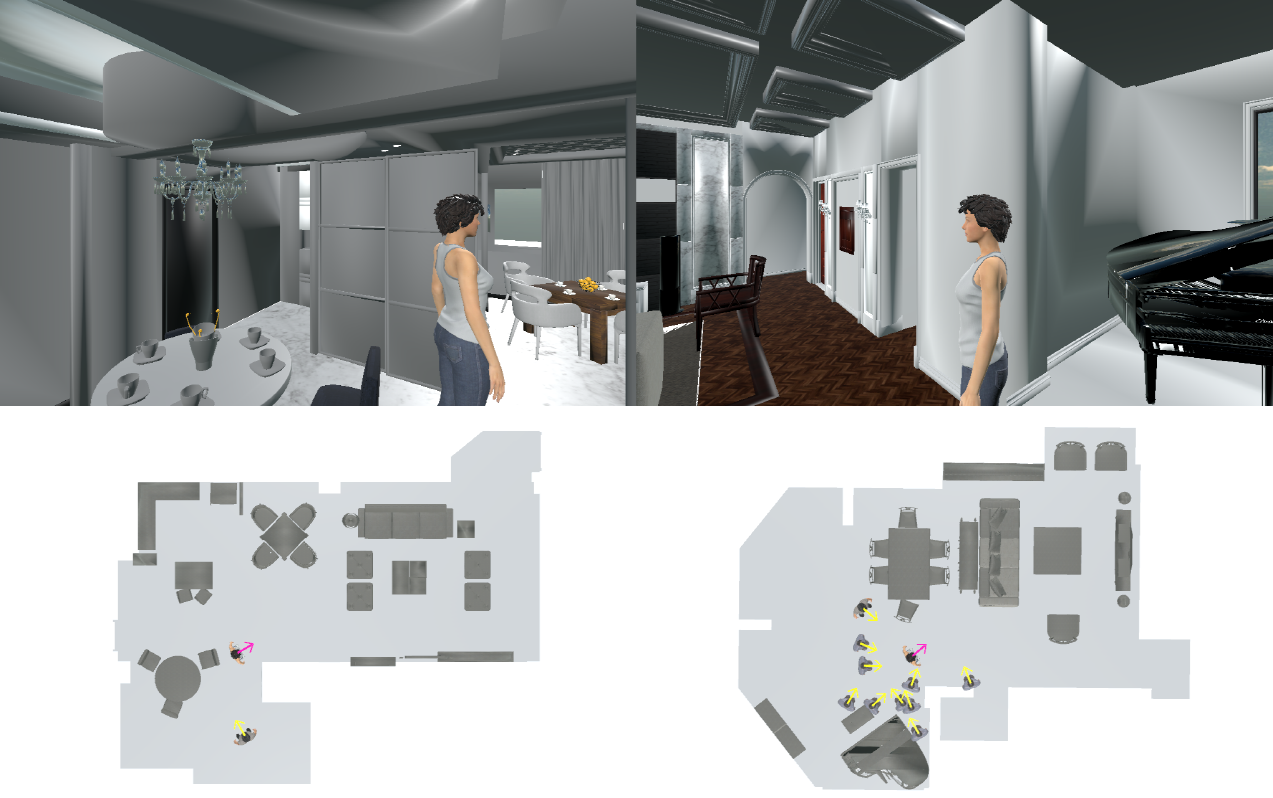}
  \caption{User survey screen. Participants can view the floor plans of Spaces A (bottom left) and B (bottom right) with placements of humans and avatars concurrently with a 2D monitor. In addition, views of Person $X$ (top left) and Avatar $X'$ (top right) are provided to the participants. Participants select placement of Avatar $X'$ corresponding to the placement of Person $X$. In the bottom right, placements of Avatar $X'$ answered by 10 participants are drawn overlapping. Yellow and purple gaze lines indicate $X$ and $Y$, respectively.}
  \label{fig:UserStudyScreen}  
\end{figure*}

\subsection{Character Placement and Action in a Scene}

To solve our problem, it is necessary to represent a human's usage of space and actions in relation to the nearby people and surrounding environment, which has been receiving growing interest in computer graphics research.

From a single indoor image, Gupta et al. \cite{gupta20113d} estimated both the 3D geometries and available human poses in the scene by simply matching a set of poses to the 3D geometries. Savva et al. \cite{Savva:2014} used observations of actual humans' interactions with objects to learn to predict the likelihood of actions in 3D scenes. They subsequently developed human-centric representations of interactions that link attributes of the human pose with the geometry and layout of the objects, by which one can generate both human poses and furniture layouts corresponding to a desired action input \cite{savva2016pigraphs}. Recently, Li et al. \cite{li2019putting} proposed a 3D pose generative model in indoor environments using the databases of semantic information from 2D images and geometric information from 3D models.

On the object interaction level, researchers have developed various methods to represent and estimate object affordance \cite{Gibson1977} or functionality, e.g., \cite{Pirk:2017:UEO,Hu:2016:LOF,Zhao:2017:IBS}.
Among them, Hu et al. \cite{hu2015interaction} proposed a geometric description of the functionality of a 3D object in the context of a given scene, derived from interactions between objects. Kim et al. \cite{Kim:2014:SHS} developed a method that predicts corresponding human poses for 3D shapes by identifying contactable local shapes in the 3D shapes and searching for admissible human poses.

Interpersonal distance is another important factor in human placement.
Proxemics \cite{hall1966hidden} is a study of humans' behavior in space use according to the egocentric distance, and a number of studies reported that proxemics theory remains valid between a person and a virtual avatar \cite{GUYE-VUILLEME:1999,BAILENSON:2003,Wilcox:2006:PSV}. In addition to interpersonal distance, Pedica and Vilhj{\'a}lmsson \cite{Pedica:2010:spontaneous}, inspired by theories on human territoriality \cite{Scheflen:1976:human}, emphasized the importance of the agents' orientation on modeling group dynamics in social interaction.

In this paper, we do not explicitly predict available actions in a scene or apply proxemics theory for avatar placement. Rather, we represent the avatar's relation to the scene and the person with low level geometric or categorical features, which reflect such high level semantics implicitly, and train a neural network to predict proper placements based on the low level features.

Planning and generating a motion by taking multiple factors into account within a scene is a complex task, and has not been researched extensively in computer graphics. One notable recent work of this kind is \cite{7127016}, which synthesized a whole-body motion that includes locomotion, body positioning, action execution and gaze behavior for generic demonstration tasks in an indoor scene with furniture. To generate a plausible motion, they considered various conditions including visibility constraints, locomotion accessibility, and action feasibility among obstacles.

\section{Method}
\label{sec:method}
A central component for finding the optimal placement of an avatar is the similarity function that estimates the similarities of the placements of the avatar and the person from the telepresence perspective. For this, we train a neural network that outputs the dissimilarity of two feature vectors that represent the semantics of the placement. As the similarity depends on human perception, we first perform a user survey that collects placement data preferred by people for a number of different configurations of space and human placements (Sec. \ref{sec:user_survey}). By examining the collected data, we identify a set of features that are relevant to the telepresence (Sec. \ref{sec:feature_modeling}). The neural network is trained to learn the similarity of the placements using the triplet loss architecture (Sec. \ref{sec:similarity_learning}). Once the similarity estimator is obtained, we find the optimal placement using a sampling-based searching scheme (Sec. \ref{sec:avatar_placement}). 

\subsection{Data Acquisition by User Survey} 
\label{sec:user_survey}
We performed a user survey that collected the preferred avatar location corresponding to a user location in various scenarios. For this, we first prepared 24 pairs of house models and generated a total of 864 questions. The detailed procedure was as follows. We selected 4 indoor 3D models available from Google Sketch Up that are adequately spacious with a reasonable number of furniture items and functional areas, such as spaces for cooking, studying, and resting. We then doubled the number of models by changing furniture arrangements and interior design (e.g., props). From 28 total pairs of 8 spaces, we excluded 4 pairs with their own variations to compose 24 space pairs. To learn the user preferred avatar placements in different indoor environments, for each space pair, we generated 36 questions by placing Person $X$, Person $Y$, and Avatar $Y'$ in different positions and orientations. We generated questions on each space so as to have a wide variety of distances and orientations between the person and avatar, interactions with objects (watching TV, looking out a window, etc.), pose (sitting or standing), and functional areas (kitchen, living room and free space). As a result, a total of 24 $\times$ 36 = 864 questions were generated.

For each question, we asked participants to place Avatar $X'$ at their preferred locations. Participants were provided with both a 2D floor plan of each space and egocentric views of the person and avatar as shown in Figure \ref{fig:UserStudyScreen}. They could explore the space by translating and rotating the avatar with a mouse. During the experiment, we let users thoroughly examine the space according to the following instruction: ``Place the avatar at a location that best represents the person's placement in a remote space''. The participants were asked to place the avatar only by looking at the views of the scene and were not provided with any additional information, such as what exactly $X$ and $Y$ are looking at or what they are doing. The reason for not giving additional information was that we wanted to develop an avatar placement algorithm that did not rely on such high level contextual information.

A total of 10 responses were obtained for each question, making total user response 864 $\times$ 10 = 8640. A total of 210 people participated in the survey and each participant answered at least 36 questions while 30 participants answered additional 36 questions with different house pair. The survey results will be publicly available online.\footnote{https://github.com/leonyoon/Avatar-Placement-User-Survey}
\newline

\begin{figure}[t]
\includegraphics[width=1\columnwidth]{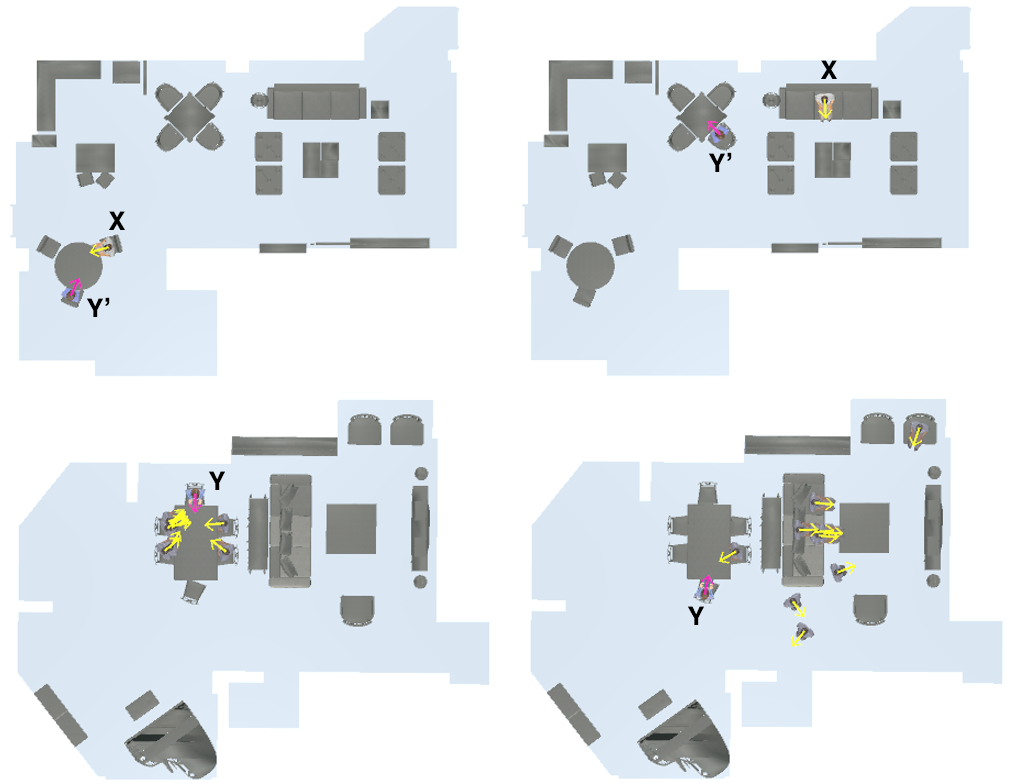}
  \caption{User survey examples. Left: When the Person $X$ and avatar $Y'$ seem to interact, users placed the Avatar $X'$ at similar locations. Right: When the Person $X$ and Avatar $Y'$ act independently, user's placements of Avatar $X'$ show higher variance. }
  \label{fig:UserStudySamples}  
\end{figure}

\begin{figure}[t]
\centering
\includegraphics[width=0.8\columnwidth]{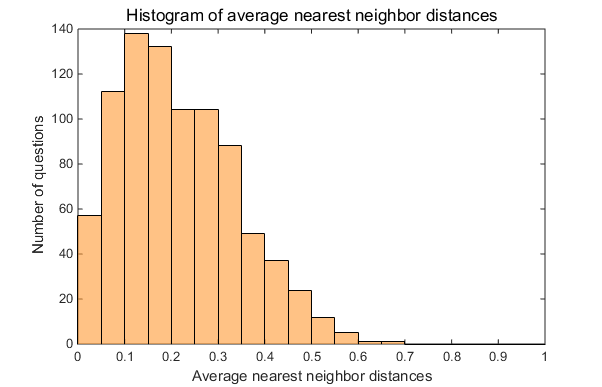}
  \caption{Histogram of the average nearest neighbor distances of the user placements in each user survey question.}
  \label{fig:UserStudyNNDistance}  
\end{figure}

\textit{Observation.} 
Figure \ref{fig:UserStudySamples} shows some typical samples from the user survey. When the two people seem to look at each other or look at an object together, participants placed avatars at places that allowed for the same action, and as a result, the participants' answers were similar. On the other hand, participants' answers varied when the configurations of two spaces were significantly different due to the separated placements of the person or the avatar, different layouts of space, or lack of a certain category of furniture. To understand the clustering/dispersion pattern of the user data, we calculated the histogram of the average nearest neighbor distances of the user placements in each question as shown in Figure \ref{fig:UserStudyNNDistance}. The nearest neighbor of each user placement was identified to be the closest one among the remaining nine user placements for the same question, with the distance defined as the weighted sum of position and angle differences. The average distances between a user placement and its nearest neighbor were used for the histogram analysis. The histogram shows the pattern of unimodal truncated normal distribution, which shows a continuously varying degree of clustering tendency of the user survey data. 

\begin{figure}[t]
\centering
\includegraphics[width=0.9\columnwidth]{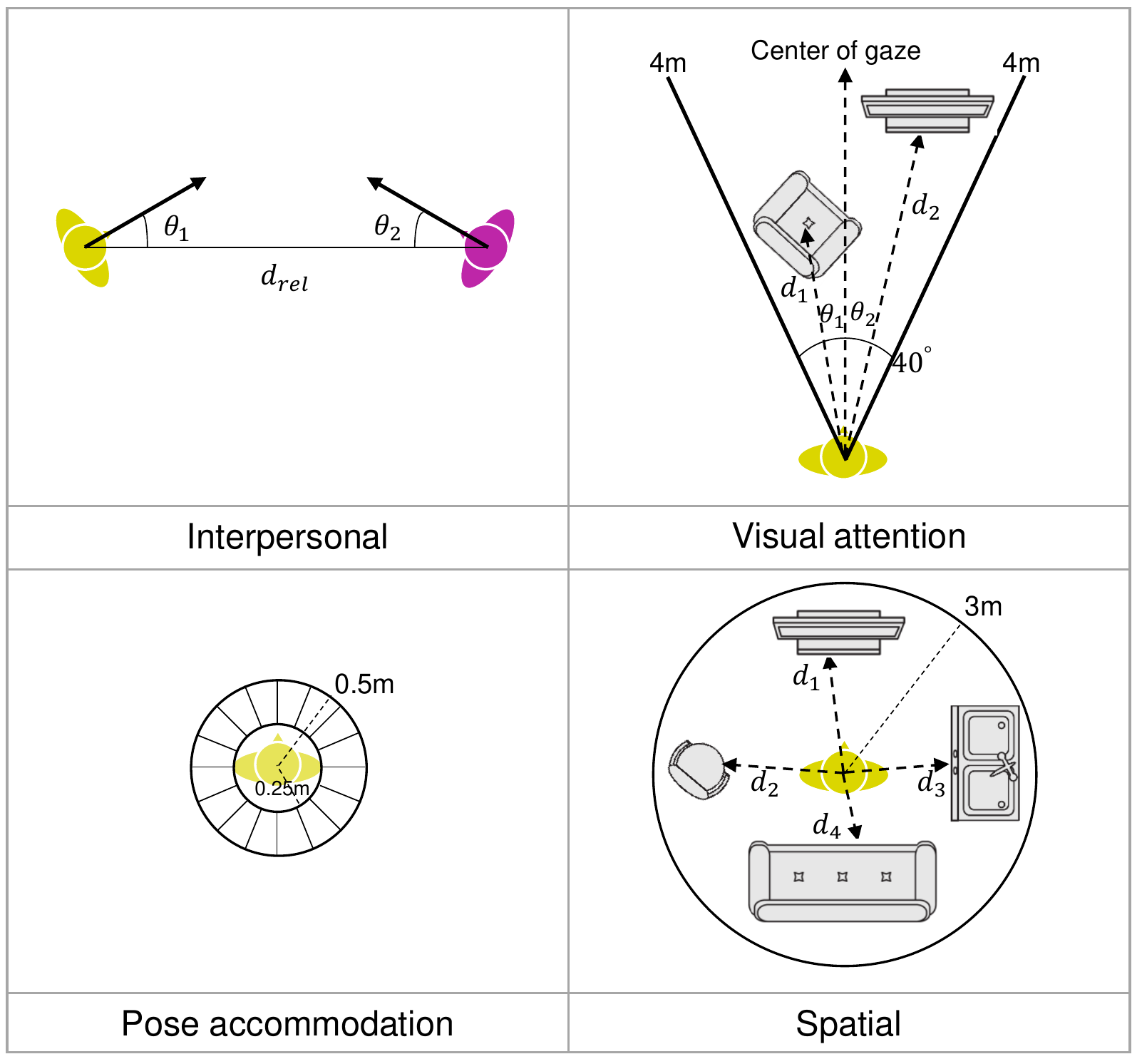}
  \caption{Features representing placement semantics.}
  \label{fig:Features}  
\end{figure}

\subsection{Feature Modeling}
\label{sec:feature_modeling}
To determine the placement of Avatar $X'$ to correspond to that of Person $X$, we need to define features that represent the meaning of the person's placement and preserve the features when the avatar is placed. After reviewing the results of the user survey, we identified the following low level features to represent the semantics of a person's placement in an indoor environment (Figure \ref{fig:Features}).

\subsection*{Interaction features} 
If a person interacts with other people or objects, and the spatial relationship between the person and an interacting entity is an important feature to be preserved in the telepresence. Distance between two people may reflect their intimacy according to Hall's proxemics \cite{hall1966hidden}. The orientation of a person's head-forward direction is also important because the visibility of the other person or an object also strongly affects the interaction. We model interaction features in two categories. One is the interpersonal relationship between a person and an avatar of the other party, which is represented with the distance and angle between the two, i.e., $x_{ip}=\{d_{rel}, \theta_{1}, \theta_2 \}$, where $d_{rel}$ is the distance between the two, $\theta_{1,2}$ is the angle difference (thus, always non-negative) from the frontal direction to the location of the other party\footnote{We chose to use the angle difference to reflect the behavior that some survey participants arranged the avatar in a direction that is reversed from the original interpersonal angle as can be seen in  Fig.~\ref{fig:UserStudySamples} (left). However, this choice allows asymmetric arrangement between the user and avatar, which may induce confusion in non-verbal communication if the user's pose is directly applied to his/her avatar (e.g., $X'$ can be on the left of $Y$ while $X$ is on the right of $Y'$. If $Y$ looks to the left to see $X'$, $Y'$ will look away from $X$). Therefore, the upper body and gaze motion should be appropriately modified by some motion retargeting algorithm, which is not investigated in this study. Replacing $\theta_{1}$ and $\theta_2$ with angles will reduce the asymmetric placement.}. The other category is the visual attention feature. In our experiment, we suppose that a person can attend to 12 object categories (sofa, chair, table, TV, air-conditioner, refrigerator, sink, lamp, piano, cabinet, shelf, window). Rather than specifying a single object that a person focuses on, which is difficult to recognize in a real situation, we assume that all objects within a narrow visual field (40 degrees) within a certain distance are candidate objects of visual attention, and nearer objects from the position of person and the center of gaze have a higher visual attention value than farther objects. Thus, the $i$-th element of visual attention feature $x_{va}$ is defined as:
\begin{equation}
x_{va,i} = \sum_j (\bar{d}_{va} - d_{i,j})(\bar\theta_{va} - \theta_{i,j}),
\end{equation}
where $d_{i,j}$ and $\theta_{i,j}$ are the distance and angle of the $j$-th object in the $i$-th category ($i = 1 \cdots 12$), and 
$\bar{d}_{va}$ and $\bar\theta_{va}$ are the maximum distance and angle (4 meters and 40 degrees in our experiment), respectively. 

Note that, unlike objects, the placement of the other party is always considered for the avatar placement through the interpersonal feature $x_{ip}$ regardless of its visibility. The rationale of this choice is that even if the two are not interacting directly, we assume that people constantly keep the position of the other party in mind. Once trained, our optimizer will give a proper importance to this feature depending on the situation.

\subsection*{Pose accommodation features} 
The pose of a person, such as sitting or standing, is an important factor that characterizes human action, and thus it would be ideal if the avatar is placed at a location that accommodates the user's pose. The pose accommodation is represented with the height field around a location because typical furniture items that accommodate human poses (e.g., floor, chair, bed) have different height fields. Specifically, we divide a circular area with a radius of 0.5 m around a person into a center, which is a circle with a radius of 0.25 m, and 16 surrounding sectors in the outer ring. The pose accommodation feature $x_{pa}$ is then represented as the average height of each subdivided area (thus forming a 17-dimensional vector). 
In addition, we add a binary feature value $x_{ss}$ to indicate whether a person or avatar is sitting or standing. In our experiment, $x_{ss}$ is set as \texttt{sit} if a person or avatar is positioned on sittable furniture categories (sofa and chair), and otherwise \texttt{stand}.

Note that the pose accommodation features not only characterize the possible poses at a given placement but also define the spatial relation between a person (or avatar) and nearby furniture items, e.g., standing beside or in front of a table.

\subsection*{Spatial features} 
The purpose of our spatial feature is to characterize the functional characteristic of the surrounding space, and the distribution of furniture is an important factor for this. To model the spatial features, we take a rather simple approach that sums up the distances of furniture items in same category within a certain distance.
Thus, the $i$-th element of spatial feature $x_{sp}$ is defined as:
\begin{equation}
x_{sp,i} = \sum_j \bar{d}_{sp} - d_{i,j},
\end{equation}
where $\bar{d}_{sp}$ is the maximum distance (3 meters in our experiment) of the spatial feature range and $d_{i,j}$ is the distance of the $j$-th object in the $i$-th category ($i = 1 \cdots 12$).

\hfill

In total, the feature vector of a placement is represented with a 45 dimensional vector $x = [x_{ip}, x_{va}, x_{pa}, x_{ss}, x_{sp}]$.

\begin{figure}
\includegraphics[width=1\columnwidth]{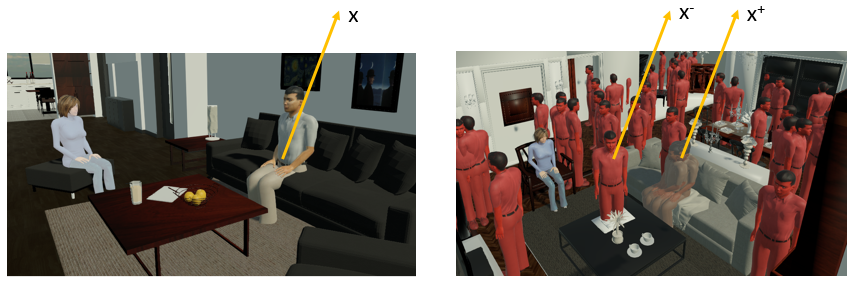}
  \caption{Left: Placement of Person $X$. Right: User selected similar placements (brown) and generated less similar placements (red) of Avatar $X'$. }
  \label{fig:PositiveNegativeData}  
\end{figure}

\subsection{Similarity Learning}
\label{sec:similarity_learning}

Until now we have obtained some samples from the user survey and defined the feature vector that characterizes a placement in an indoor scene. Let us denote the feature of the placement of Person X by $x^0$ and that of Avatar $X'$ obtained from the user survey by $x^+$. For each sample of $x^0_i$, we have obtained $x^+_{i,j},~(i=1\cdots 864,~j=1\cdots 10)$. 

It is to be noted that the user-selected placements indicate that they are better than other placements, but not the absolute answers. That is, a user sample $x^+$ might not have been selected if there had been better positions. Therefore, this problem needs to be addressed in terms of ranking between data, i.e., by training a similarity function $d(\cdot,\cdot)$ such that it can judge $d(x^0,x^+) < d(x^0,x^-)$ where $x^-$ is a feature from the placements not selected by the users. 

A basic approach for this is to obtain the similarity function in bilinear form, i.e., $d(x^0,x) = (x^0-x)^T W (x^0-x)$. Many techniques have been developed for the similarity learning or the distance metric learning in the bilinear form \cite{weinberger2006distance, davis2007information}, among which some support learning relative similarity as in our problem \cite{mcfee2010metric}. This approach would work fine if the difference in data, $x^0-x$, provided enough information for ranking. Unfortunately, this is not the case for our problem because the importance of subfeatures varies according to the configuration of $x^0$. In order to reflect the non-linear characteristic of data similarity in our problem, we train a deep neural network that learns the similarity between two placement features, as will be introduced next.

\subsubsection{Dataset Preparation} 
We obtained similar pairs ($x^0$,$x^+$) from the user survey, but we also need less similar (will be called ``dissimilar'' hereafter) pairs ($x^0$,$x^-$) for the training. To generate dissimilar features $x^-$ given $x^0$ in a scene, we generated random placements from the scene and computed corresponding features to the placements (Figure \ref{fig:PositiveNegativeData}). {\blue A random placement sample that was not too close to one of the user-selected placements with respect to  distance (1 meter), angle (36 degree), and pose (10\% of normalized value) was collected over the whole indoor space for 100 samples,} and we added 10 samples near user-selected placements to make challenging dissimilar data. Thus, for each $x^0$, we have 10 positive data $x^+$ and 110 negative data $x^-$, which makes a total of 1100 tuples $(x^0, x^+, x^-)$ used for training.

\begin{figure*}
    \subfigure[A typical Triplet Matchnet structure.]{
        \centering
        \includegraphics[width=0.25\textwidth]{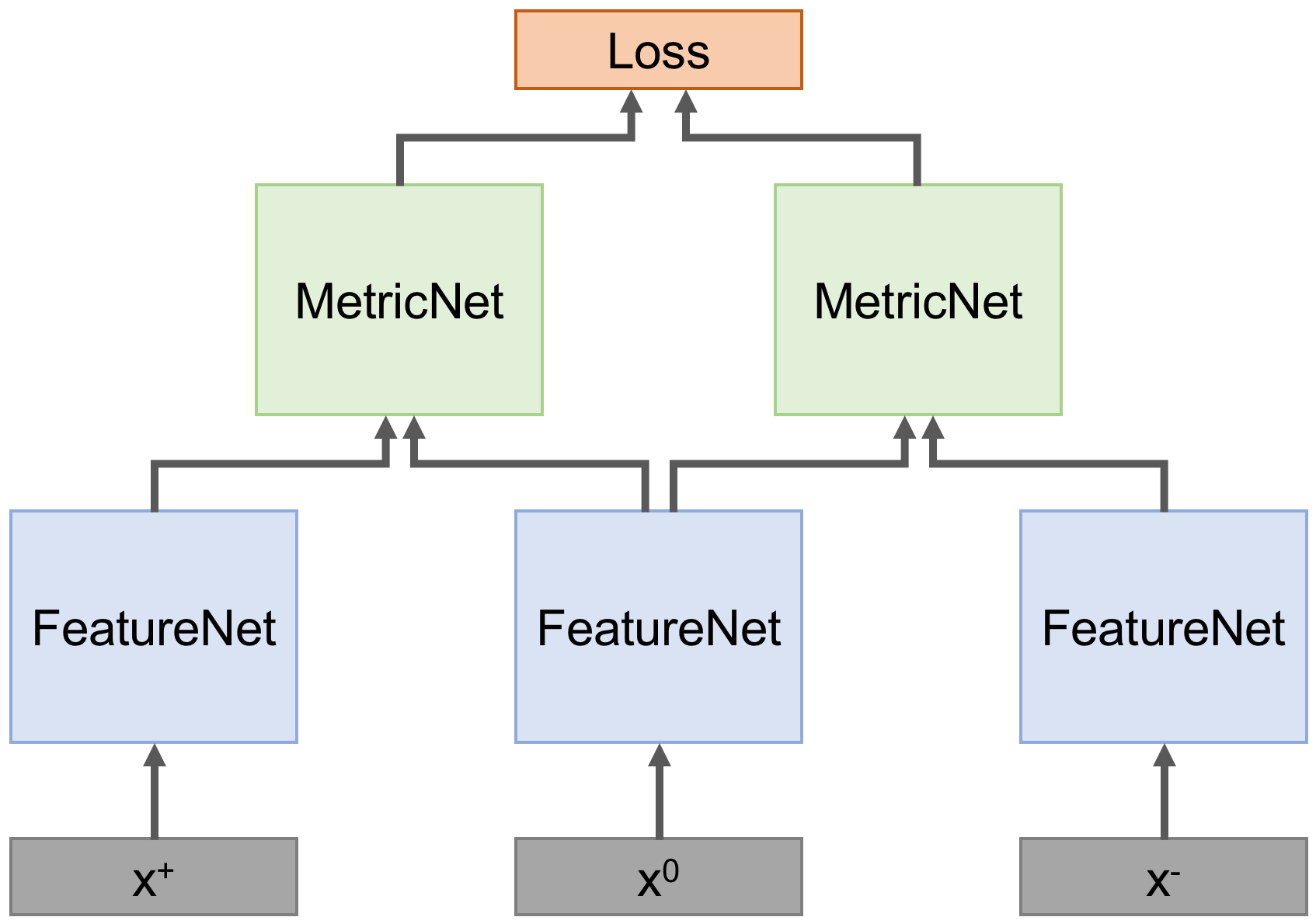}
        \label{fig:TripletMatchNet}
    }
    \subfigure[Our modified Triplet Matchnet structure.]{
        \centering
        \includegraphics[width=0.3\textwidth]{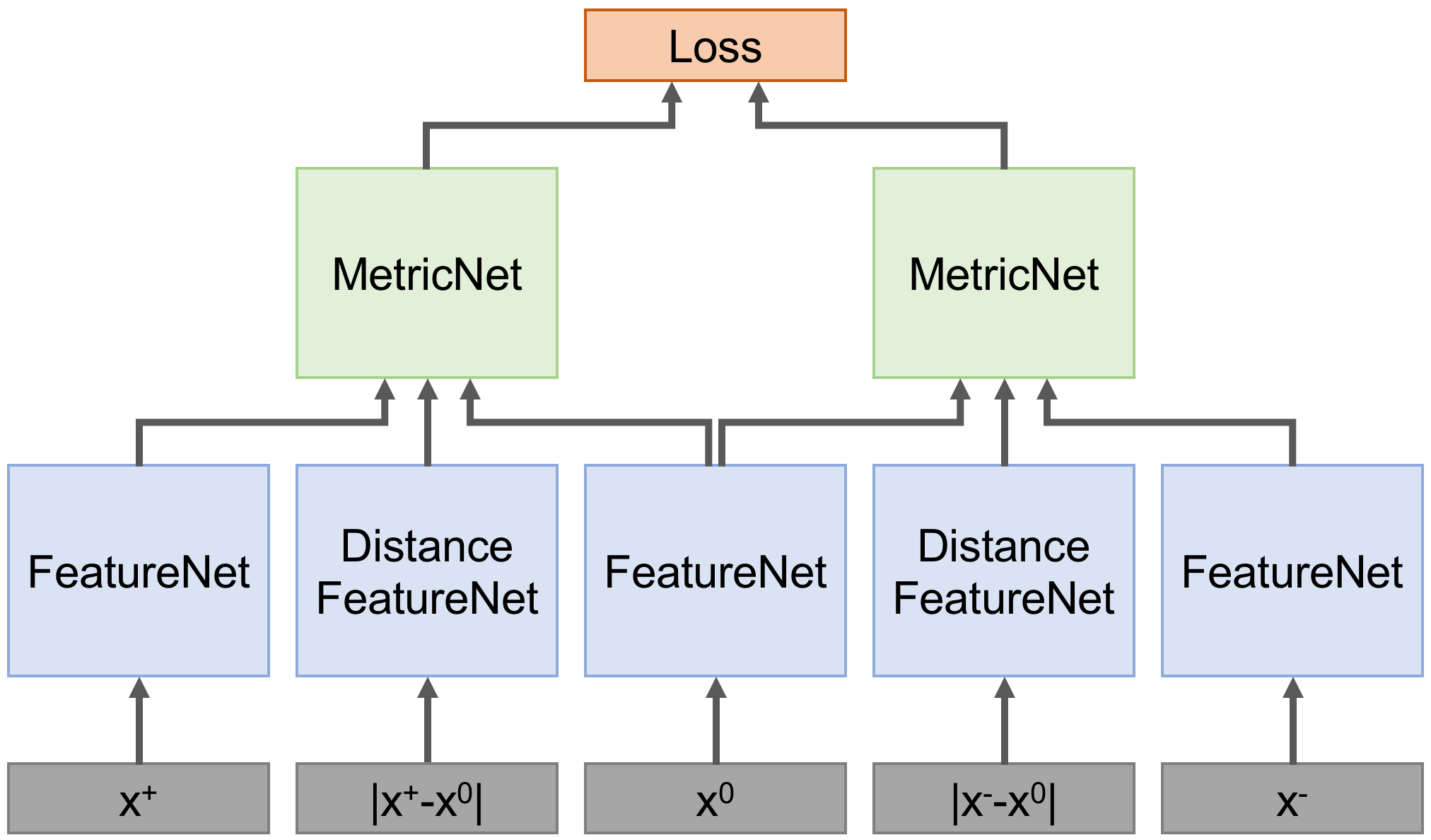}
        \label{fig:Quintuple_MatchNet}
    }
    \subfigure[BBM (left) and SFPM (right) modules in our neural network.]{
        \centering
        \includegraphics[width=0.4\textwidth]{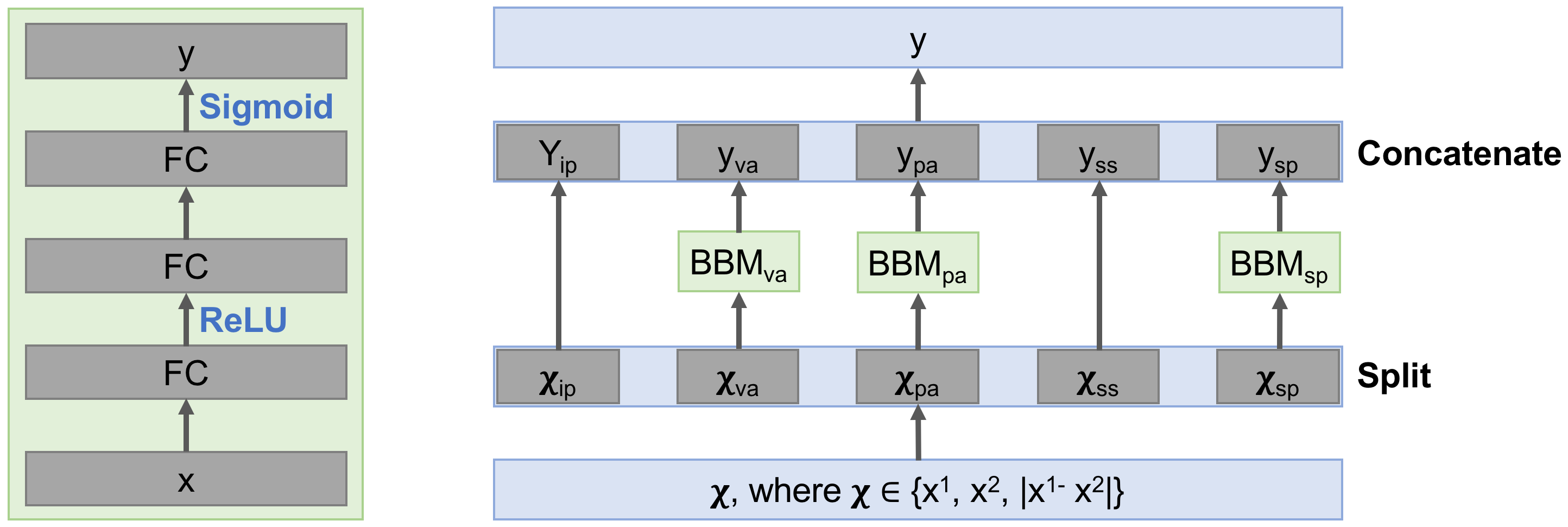}
        \label{fig:BBM_SFPM}
    }
    \caption{Network structure of our placement similarity predictor.}
    \label{fig:network_structure}      
\end{figure*}

\subsubsection{Learning} 
We develop a neural network that learns the non-linear characteristics of the dissimilarity (or distance) between two placements. The neural network is trained with triplet loss framework \cite{hoffer2015deep}, which learns the dissimilarity between input features in a supervised way (Figure \ref{fig:TripletMatchNet}). {\blue This framework for learning a ranking from relative similarity has been proved to be effective not only for information retrieval \cite{wang2014learning} and recommendation system \cite{lei2016comparative}, but also for general problems including classification and clustering that use a distance metric \cite{hoffer2015deep}.} In the training phase, the model receives datasets of three types of inputs $\{x\}$: $x^0$, $x^+$, and $x^-$, then the network learns the dissimilarity between input features to predict that $x^0$ is more similar to  $x^+$ than $x^-$. 
Specifically, the network is optimized to push the distance $d^+$ between the similar pair ($x^0, x^+$)  to be low (zero), and the distance $d^-$ between the dissimilar pair ($x^0, x^-$) to be high (one). At the same time, the comparison between the two distances are utilized so that $d^+$ is less than $d^-$. Following \cite{lu2017deep}, in our work the two optimization goals are achieved with the loss functions $\phi (\{x\})$ and $\psi (\{x\})$:
\begin{align}
	\label{eq:LossFunction1}
\phi (\{x\}) &= - \frac{1}{|\{x\}|} \sum_{x \in \{x\}} \{\log(1-d^+) + \log(d^-)\}, \\
	\label{eq:LossFunction2}
\psi (\{x\}) &= \frac{1}{|\{x\}|} \sum_{x \in \{x\}} \max\{0, d^+ - d^-\},
\end{align}
where the total loss function is defined as $\phi (\{x\}) + \psi (\{x\})$. 

\subsubsection{Our Network Structure} 
Typical triplet network models \cite{lu2017deep, qi2017audio} consist of FeatureNet for feature abstraction and MetricNet for the dissimilarity estimation. 
We slightly expand the triplet structure to provide an additional hint for the neural network to estimate dissimilarity better for our problem.
In addition to the triplet inputs, our model explicitly computes the differences between two features $|x^0-x^\pm|$ and uses them as additional inputs. 
Figure \ref{fig:Quintuple_MatchNet} describes the overall structure of the network model. The explicit distance inputs are passed through another FeatureNet, dubbed Distance FeatureNet. 

Our neural network model consists of two basic modules: a black box module (BBM) and a subfeature processing module (SFPM) as shown in Figure \ref{fig:BBM_SFPM}.The SFPM receives one of the input feature sets, $\chi \in \{x^0, x^+, x^-, |x^0-x^+|, |x^0-x^-|\}$, and splits the input into the same type of subfeatures, $x_{ip}$, $x_{pa}$, $x_{va}$, $x_{ss}$, and $x_{sp}$. The high dimensional subfeatures with local correlation, $x_{pa}$, $x_{va}$, and $x_{sp}$, are processed through separate BBMs. Then, the outputs of the BBMs and low dimensional subfeatures, $x_{ip}$ and $x_{ss}$, are concatenated, making the final outputs of the SFPM. Two SFPMs are used as FeatureNet and Distance FeatureNet at the bottom, and a simple BBM is used as MetricNet at the top.

\begin{table}
\small
\begin{center}
\begin{tabular}{|l|c|c|c|c|}
\hline
Model & x & $h_1$ & $h_2$ & y\\
\hline\hline
BBM$_{\text{pa}}$ & 17 & 14 & 10 & 6 \\
BBM$_{\text{sp}}$ & 12 & 10 & 8 & 6 \\
BBM$_{\text{va}}$ & 12 & 10 & 8 & 6\\
BBM FeatureNet & 45 & 38 & 30 & 22 \\
BBM MetricNet & $22 \times N$ & 44 & 44 & 1 \\
\hline
\end{tabular}
\end{center}
\caption{The number of cells in different BBMs; input cells (x), cells in subsequent layers ($h_1$ and $h_2$), and output cells (y). $N$ is two for the regular Triplet MatchNet and three for our model.}
\label{tab:BBMCellSize}
\end{table}

Our models have five types of BBMs and their layer sizes are reported in Table \ref{tab:BBMCellSize}. In our experiments, we compared the BBM with the SFPM as FeatureNet (BBM FeatureNet). Each layer of this BBM kept the same number of cells with corresponding layer cells in the SFPM. 

\subsection{Avatar Placement Algorithm}
\label{sec:avatar_placement}

We use a sampling-based optimization scheme to find the optimal placement of Avatar $X'$ given the placements of Person $X$ in a dissimilar space. 
For brevity, let us write $d( x( p_{X'} | p_Y, B), x( p_X | p_{Y'}, A) )$, the dissimilarity between the placement of Avatar $X'$ and Person $X$, as $D(p_{X'},p_{X})$. Then our optimization problem is 
\begin{equation}
    \argmin_{p_{X'}} D(p_{X'},p_{X}).
    \label{eq:placement}
\end{equation}
We solve this optimization in two stages. First, we sample each space as a 2D grid map with a grid size of 0.25 meter and for each grid we take 24 orientation samples for each 15 degrees. We compare the dissimilarity between the feature vector of Person $X$ and that of Avatar $X'$ at every sample placement to find the best sample with the lowest dissimilarity in the given grid.
Next, starting from the best sample placement we use the particle swarm optimization (PSO) to find the optimal placement of the Avatar $X'$. In our experiment, we used inertia weight w = 0.73, constriction factors $c_{1}$,  $c_{2}$ = 1.49 \cite{eberhart2000comparing}, 10 particles and 10 maximum epochs as to the PSO parameters. 

The placement algorithm must run fast to place avatars promptly in the real-time applications. To this end, we precompute space-specific features such as $x_{pa}$ and $x_{sp}$ and the 2D obstacle meshes for each space.
In the on-line process, we use multi-threading (16 threads) for operating the static placement, achieving 0.4 sec (0.3 sec for grid-level optimization, 0.1 sec for PSO) for each placement.

\begin{figure}[t]
\includegraphics[width=1\columnwidth]{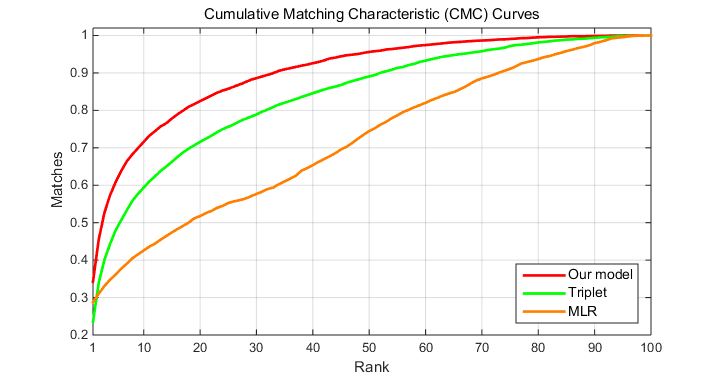}
  \caption{CMC curves for Triplet model, our model, and a linear model (MLR).}
  \label{fig:CMC}  
\end{figure}

\begin{figure}[h]
    \centering
    \subfigure[Average Rank 1\%]{
        \centering
        \includegraphics[width=1\columnwidth]{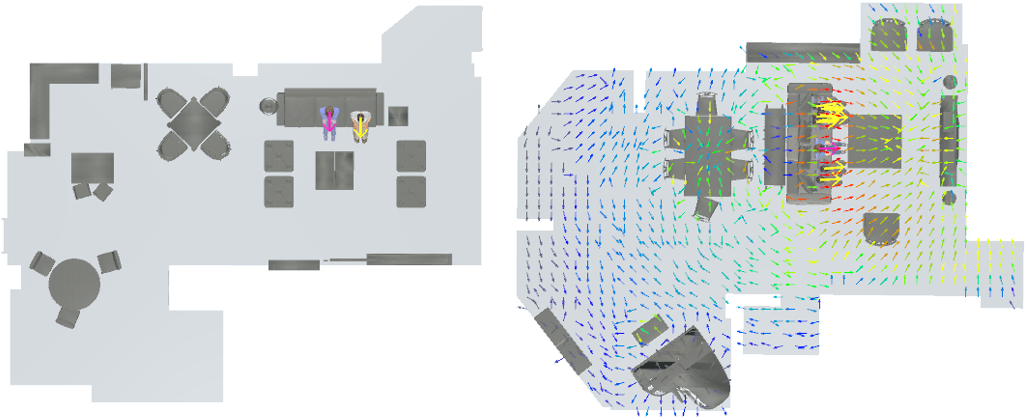}
        \label{fig:Rank1}
    }
    \subfigure[Average Rank 5\%]{
        \centering
        \includegraphics[width=1\columnwidth]{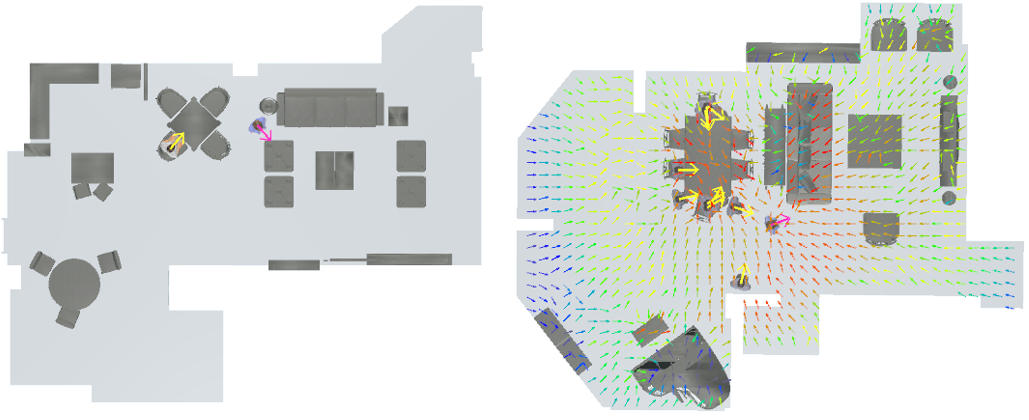}
        \label{fig:Rank5}
    }
    \subfigure[Average Rank 35\%]{
        \centering
        \includegraphics[width=1\columnwidth]{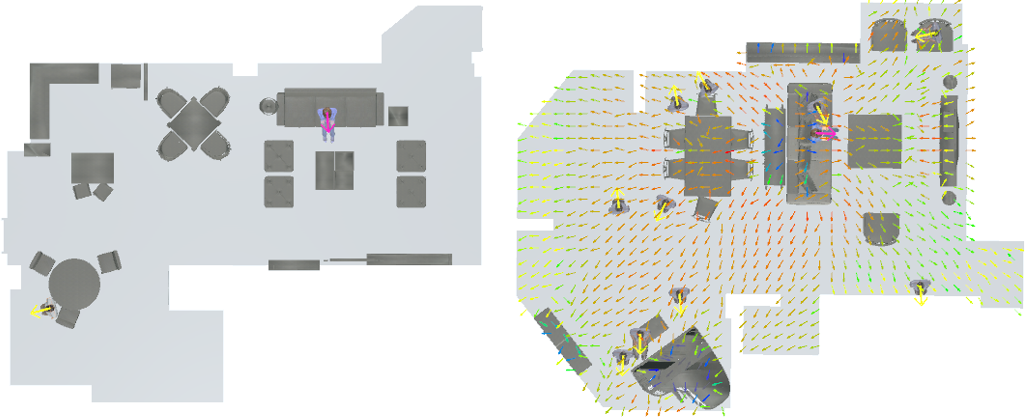}
        \label{fig:Rank35}
    }
    \caption{Heat maps when the average rank of user-selected placements are 1, 5, and 35\%.}
    \label{fig:heatmap}  
\end{figure}

\section{Results and analysis}
\label{sec:result}
We compare our triplet-based method with a linear ranking learning algorithm as a baseline method as well as with the variations of our own network structure with respect to the learning accuracy. For the cross-validation, data from half of all house pairs are used for training and the rest of the data are used for testing.
In addition, we develop a VR and AR telepresence prototype system and perform user study to evaluate our system from the user experience perspective.

\subsection{Similarity Learning Accuracy}

\begin{table}[t]
\small
\begin{center}
\begin{tabular}{|c|l|c|c|}
\hline
Rank & Model & Avg. Acc. & Std. Dev. \\
\hline\hline
1 & SFPM, DFN (proposed) & 96.7\% & 0.0008 \\
2 & BBM, DFN & 96.3\% & 0.0007 \\
3 & SFPM, No DFN & 95.6\% & 0.0007 \\
4 & BBM, No DFN & 92.4\% & 0.0050 \\
\hline
\end{tabular}
\end{center}
\caption{Test accuracies of proposed method and its variations.}
\label{tab:accuracy}
\end{table}

As a baseline, we use the structural SVM approach for metric learning to rank (MLR) \cite{mcfee2010metric}, a well-known method for the information retrieval. As a generalization of a multi-class SVM, the structural SVM framework defines the loss function in terms of ranking. The MLR learns a positive semi definite matrix $W$ to maximally satisfy $||x^0 - x^+||_W < ||x^0 - x^-||_W$ for the training set where $||i - j||_W = \sqrt{(i - j)^TW(i - j)}$. A major drawback is that this linear metric cannot reflect the inherent non-linear characteristic of our problem. 

We use the cumulative matching characteristic (CMC) curve to compare deep neural network variations and the structural SVM approach. For each test question, we rank 10 user-selected placements as well as all other sample placements and sort their dissimilarity values in increasing order. For normalization, we divide the rank by the total number of placements to represent it as a percentage. For example, if the rank of one user data is 10 out of 2,000 possible sample placements, we divide 10 by 2,000 and apply a ceiling function to represent it as a percentile rank 1. Figure \ref{fig:CMC} shows that our model surpasses the baseline method by a large margin and is superior to the baseline triplet network method used in \cite{lu2017deep}.

Figure \ref{fig:heatmap} shows three examples of different matching characteristics. On the input of Person $X$ (yellow) in the left space, the colored arrows in the right indicate the similarity (red: high, blue: low) and optimal direction at each location computed by our model. Figure \ref{fig:Rank1} shows a case in which the average percentile rank of a user selected placement is 1\%. The heatmap shows that high similarity placements are concentrated at the sofa, which agrees well with user-selected placements. Figure \ref{fig:Rank5} shows a similar pattern, but the standing area near Person $Y$ receives increased similarity values due to the interpersonal relation feature. Lastly, Figure \ref{fig:Rank35} shows a case in which the user-selected placements vary widely, mostly looking at window while standing near furniture. The heatmap shows a similar tendency of giving higher values to a wider area, which produced a lower average rank of user-selected placements. We also compare our proposed network structure with its variations on test accuracy. The variations are made in two aspects: use of the SFPM for FeatureNet and Distance FeatureNet (SFPM vs. BBM), and use of the Distance FeatureNet (DFN vs. no DFN). 

Training and test sets are selected to have the same number of samples, 150 for each, and not to have the samples from the same space. As shown in Table \ref{tab:accuracy}, our network model outperforms its variations in terms of the test accuracy.

\begin{figure}[t]
    \centering
    \subfigure[House pair \#1. Installed objects: Table, TV, sink, chair, sofa, lamp, and cabinet in both houses. Shelf and window  in the right house only.]{
        \centering
        \includegraphics[width=1\columnwidth]{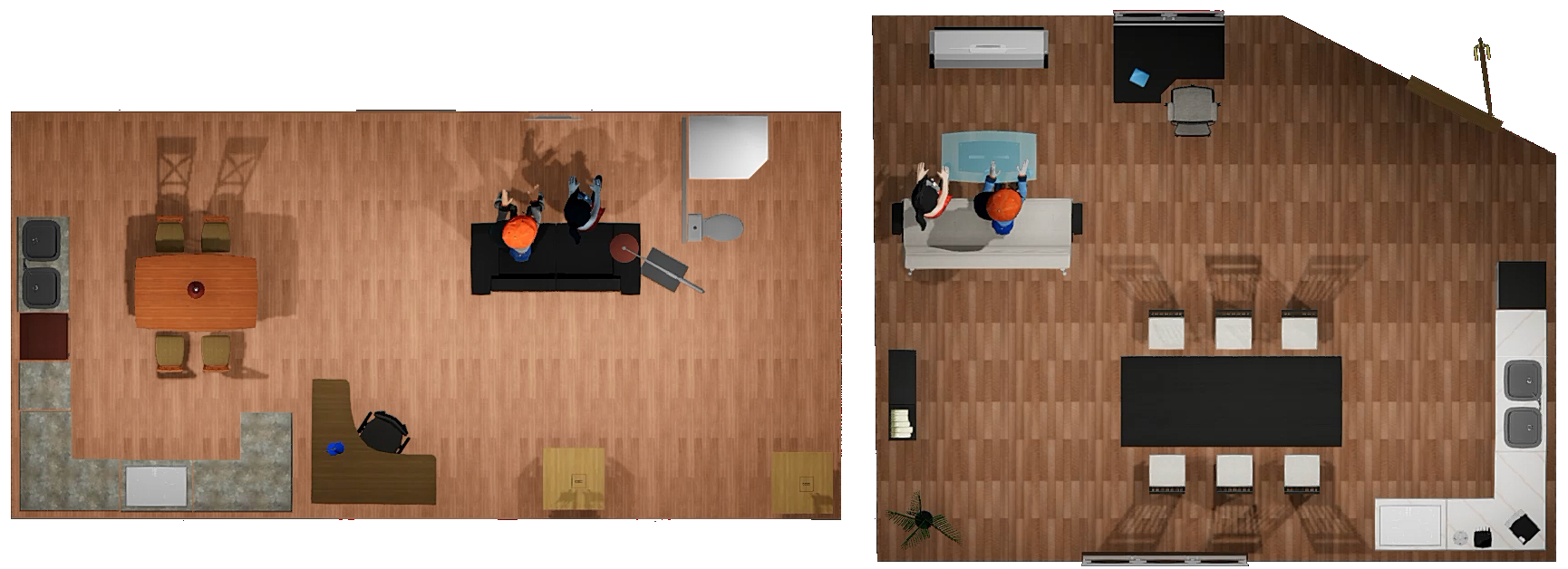}
        \label{fig:housePair1}
    }
    \subfigure[Perspective view.]{
        \centering
        \includegraphics[width=1\columnwidth]{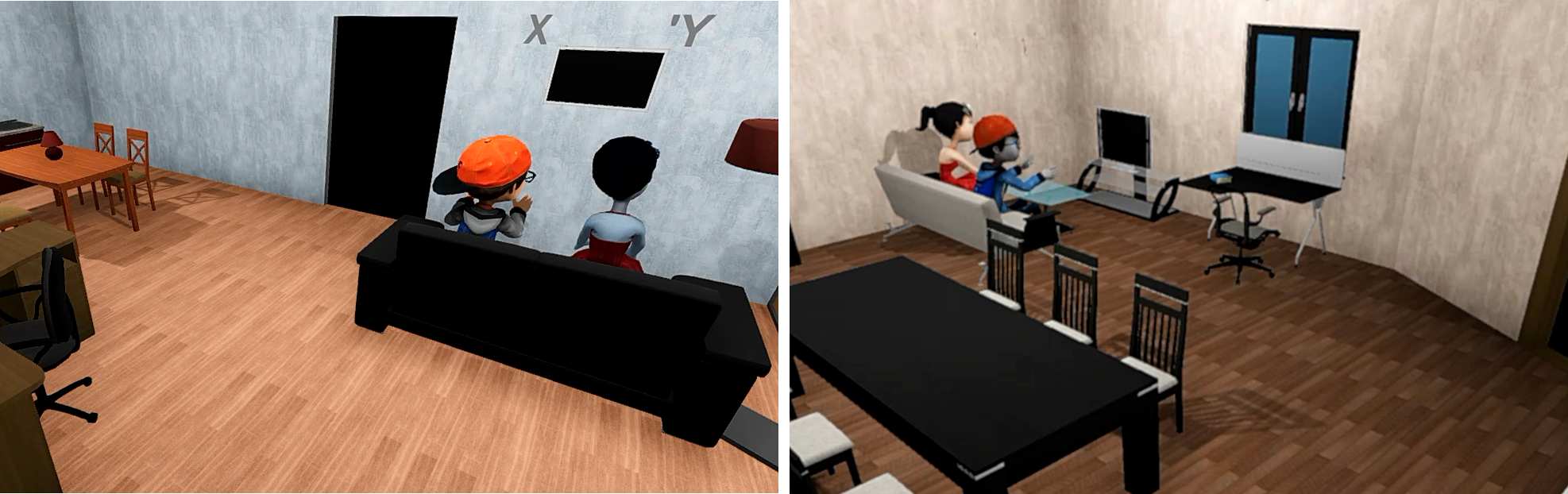}
    }   
    \caption{Joint attention case. Both users look at TV.}
    \label{fig:VRInteraction}
\end{figure}

\begin{figure}[t]
    \centering
    \subfigure[House pair \#2. Installed objects: Table, TV, chair, sofa, lamp, and cabinet in both houses. Sink, shelf, and window  in the right house only.]{
        \centering
        \includegraphics[width=1\columnwidth]{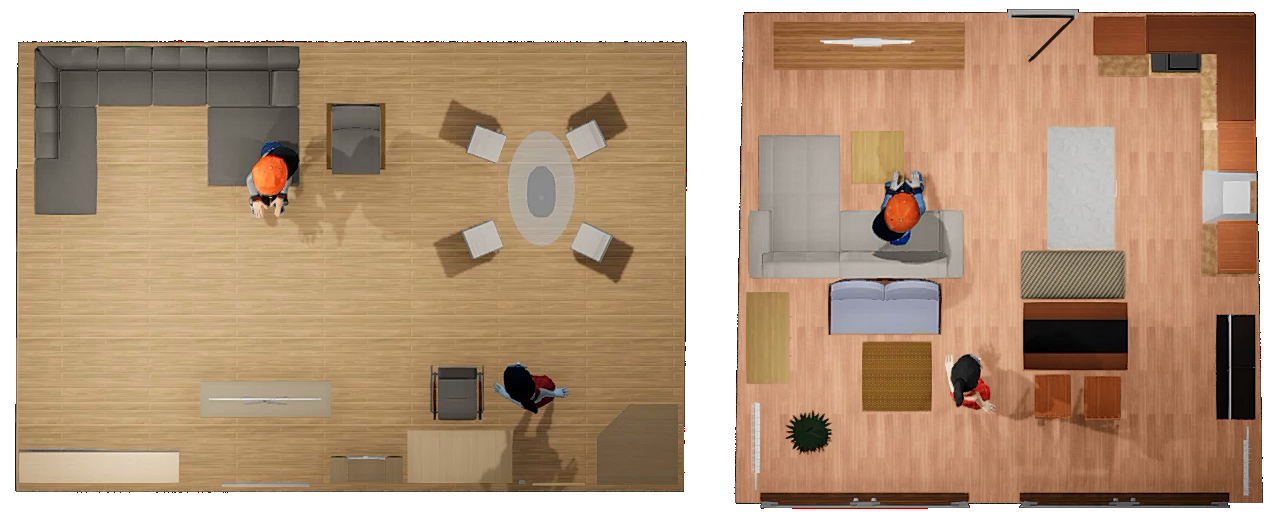}
        \label{fig:housePair2}
    }
    \subfigure[Perspective view.]{
        \centering
        \includegraphics[width=1\columnwidth]{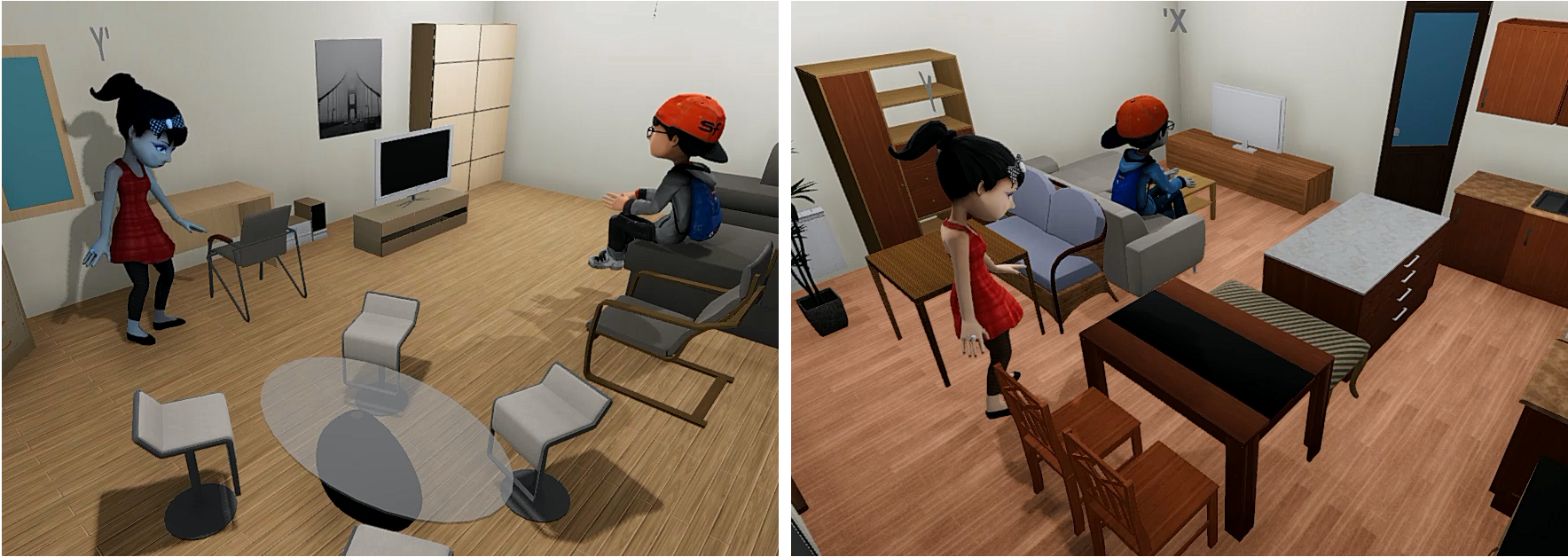}
    } 
    \caption{Individual action case. }
    \label{fig:VRSpatial}
\end{figure}

\subsection{Telepresence System Construction}
\label{sec:experiment}
To validate avatar placement algorithm in immersive VR and AR environments, we develop a prototype telepresence system with HTC Vive Pro system and Unity3D game engine, and conduct experiments. 
Our telepresence system consists of two separate stations.
In each station, one user is located in a separate indoor space and can see the avatar of the other party visiting his/her space.
For VR, the users view the virtual scene with an HMD and can freely move in an empty physical space, with stools placed where virtual chairs are located to accommodate sitting. For AR, the users view the real scene with a video see-through HMD and interact with real furniture in the space. The registration of the virtual object and the real object is obtained manually using an input device. 
A user's movement is captured with three trackers attached on the body (on the waist and feet), handheld controllers and HMD. Full body pose is obtained with inverse kinematics. Data transferred between stations through the Internet are the feature vector of person's placement as well as positions and orientations at six parts of the body. 

\subsection{VR Experiments}
\label{subsec:experiment}
While our method was developed for AR telepresence, we perform test in VR first as it is more convenient to construct various spatial configurations in VR.
For VR telepresence test, we select 4 studio-type house models from SunCG database \cite{song2016ssc} to make two pairs of houses (Figure \ref{fig:housePair1} and \ref{fig:housePair2}).
The first pair of houses are selected such that both houses have the same functional areas, kitchen, dining area, living room, and study area, but with different layouts. The second pair of houses is made to give a more challenging situation: both houses have a living room and a dining area, but a kitchen area exists only in the right house and a study area exists only in the left house. 

For each house pair, we experiment avatar placement for a number of user positions. Figure \ref{fig:VRInteraction} shows a case that our system realizes a joint attention. When a user approach an avatar of the other party and looks at the same place, his/her avatar is placed in the vicinity of the other user and face to the same object. Figure \ref{fig:VRSpatial} is a case in which two users interact with objects without regard to each other. Their avatars find their placements in appropriate locations to preserve the users' poses and their interaction with objects. Supplementary video shows more examples for VR experiments.

\subsection{Comparison with Other Approaches}
\label{subsec:comparison}

Some studies in telepresence \cite{Lehment2014} determine affine transformation between two spaces and place avatars according the transformation. This approach keeps only the interpersonal relationship in shared free space while not supporting other semantic features. Other studies \cite{Pejsa:2016, Jo:2015} are more comparable to our work, but they mainly deal with sitting affordance. To show the strength of our method that considers a wide range of features, we compare the placements by our method with those that match only a few features (e.g., sitting affordance, visual attention, and interpersonal relation) adopted in the previous work.

\begin{figure}[t]
    \centering
    \subfigure[Interaction around the table.]{
        \centering
        \includegraphics[width=0.96\columnwidth]{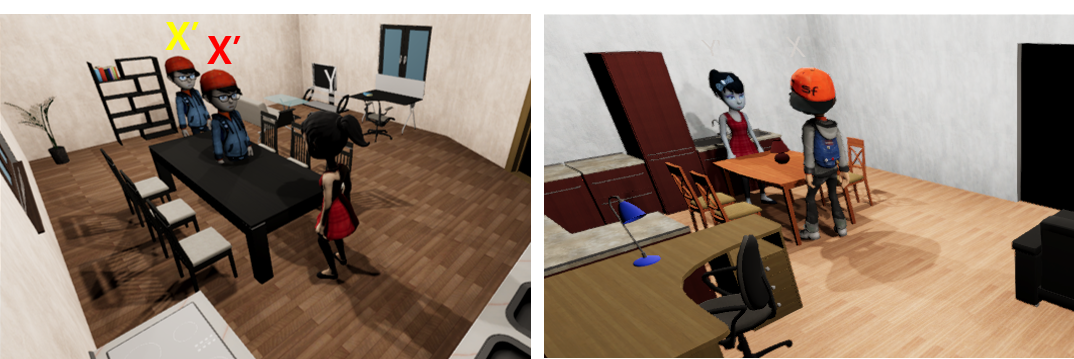}
        \label{fig:comparison1}
    }
    \subfigure[Individually sitting in dining table.]{
        \centering
        \includegraphics[width=0.96\columnwidth]{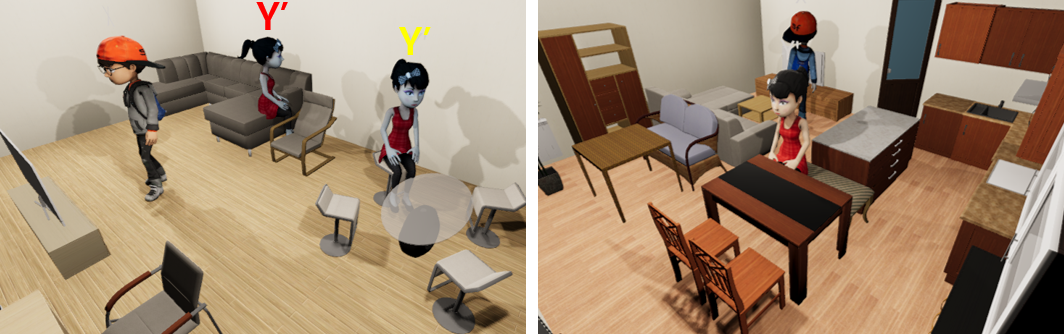}
        \label{fig:comparison2}
    }
    \subfigure[Watching TV together.]{
        \centering
        \includegraphics[width=0.96\columnwidth]{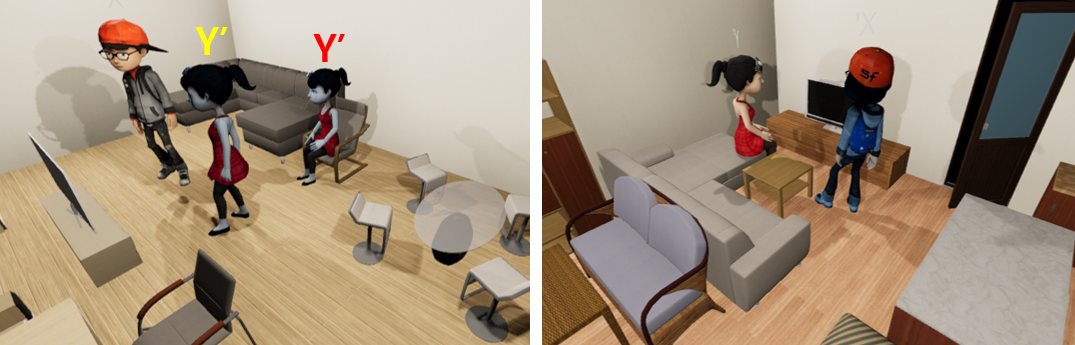}
        \label{fig:comparison3}
    }
    \caption{{{\blue Comparison between the results of our method (yellow) and the results of the avatar placement that matches only a few features (red). The right part is the original configuration around the person and the left part shows the results of the avatar placement.}}}
    \label{fig:comparison}
\end{figure}

Figure \ref{fig:comparison} shows three examples of avatar placement by our method for comparison. Figure \ref{fig:comparison1} shows that the placement of Avatar X$'$ by our method preserves the context of interaction between two people around the table by coordinating interpersonal relation, pose affordance, spatial and visual attention feature of Person X, compared to the placement of avatar (X$'$ in red) that only considers interpersonal relation. Second, Figure \ref{fig:comparison2} shows that the placement of Avatar Y$'$ by our method preserves the context of sitting in dining table. In comparison, the result obtained by matching only the interpersonal relation and sitting affordance (Y$'$ in red) does not respect this spatial characteristic. Third, our method preserves the context of watching TV together in Figure \ref{fig:comparison3}. For this case, our method keeps the visual attention and interpersonal relation over the sitting affordance (Y$'$ in red).

\begin{figure*}[t]
    \centering
    \subfigure[A local user $X$ is sitting and looking at the avatar $Y'$.]{
        \centering
        \includegraphics[width=0.9\columnwidth]{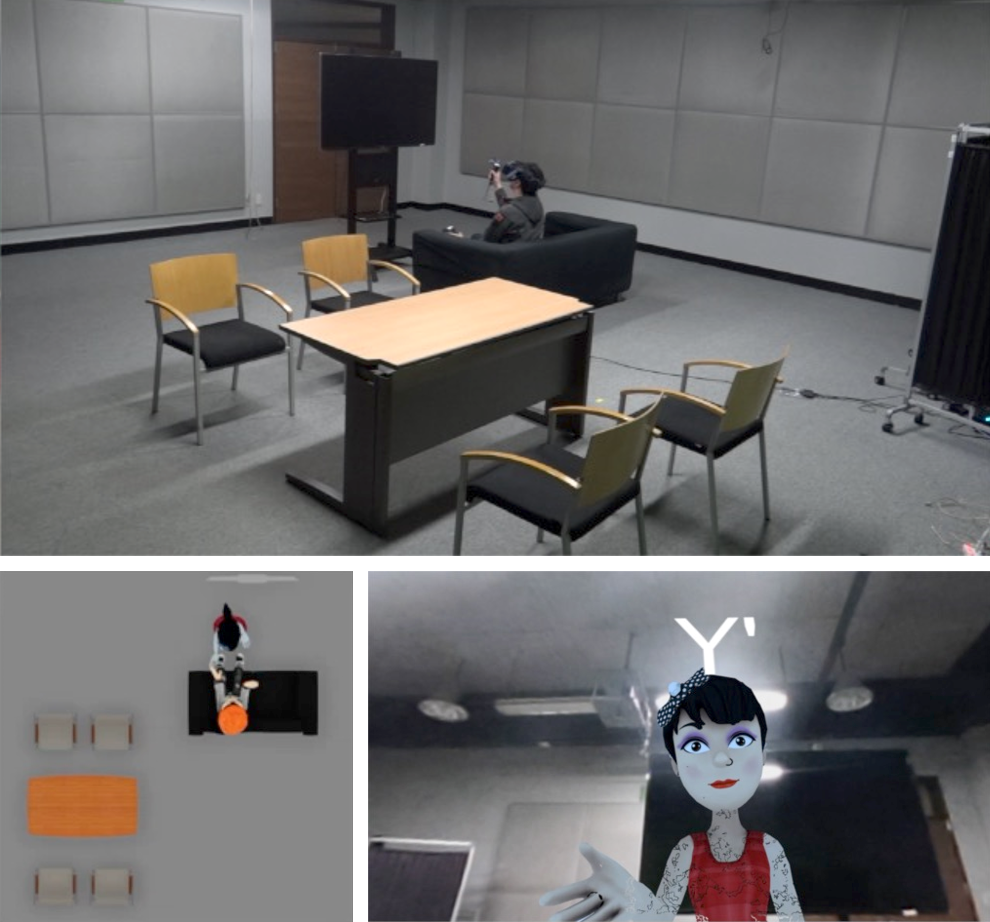}
    }
    \subfigure[A local user $Y$ is standing and looking at the avatar $X'$.]{
        \centering
        \includegraphics[width=0.9\columnwidth]{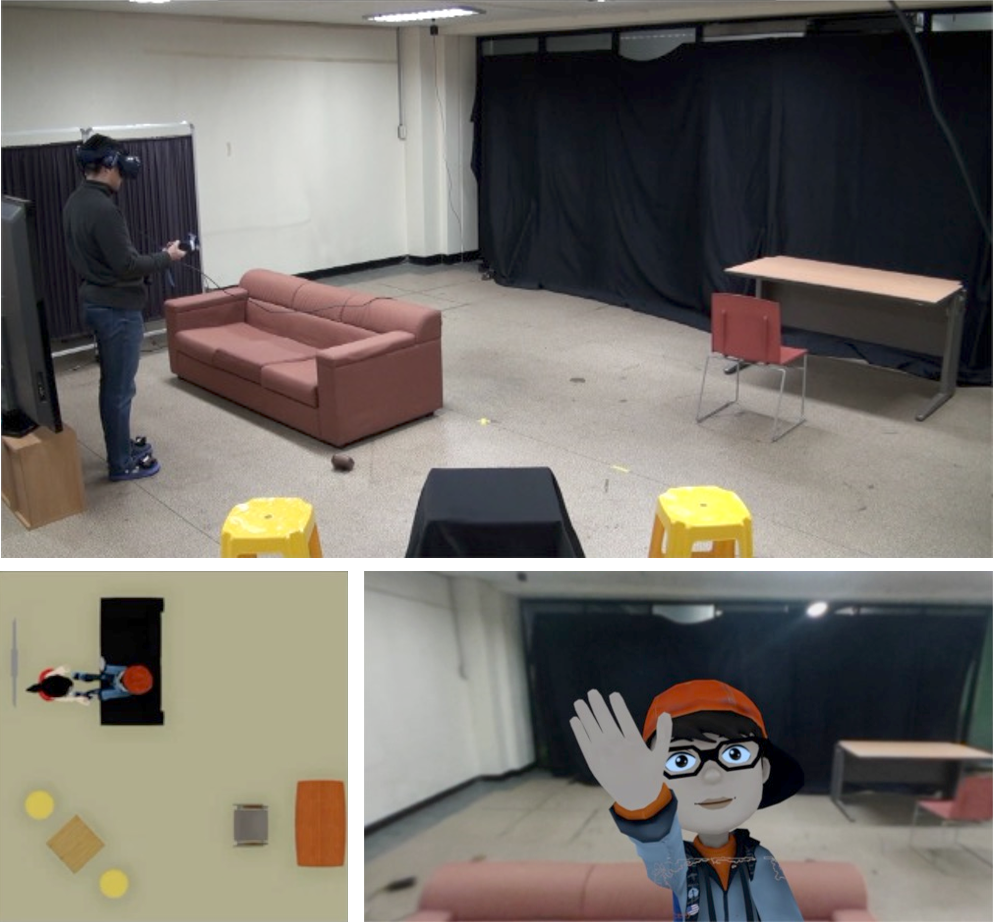}
    }    
    \caption{Interaction case in the AR experiment. In each room, perspective (top) and top (bottom left) views of the room, and the HMD view of the user (bottom right) are shown. The top views show the virtual 3D models corresponding to the real objects.}
    \label{fig:AR_interaction}
\end{figure*}

\begin{figure}[h]
 \centering
 \subfigure{
 \includegraphics[width=\columnwidth]{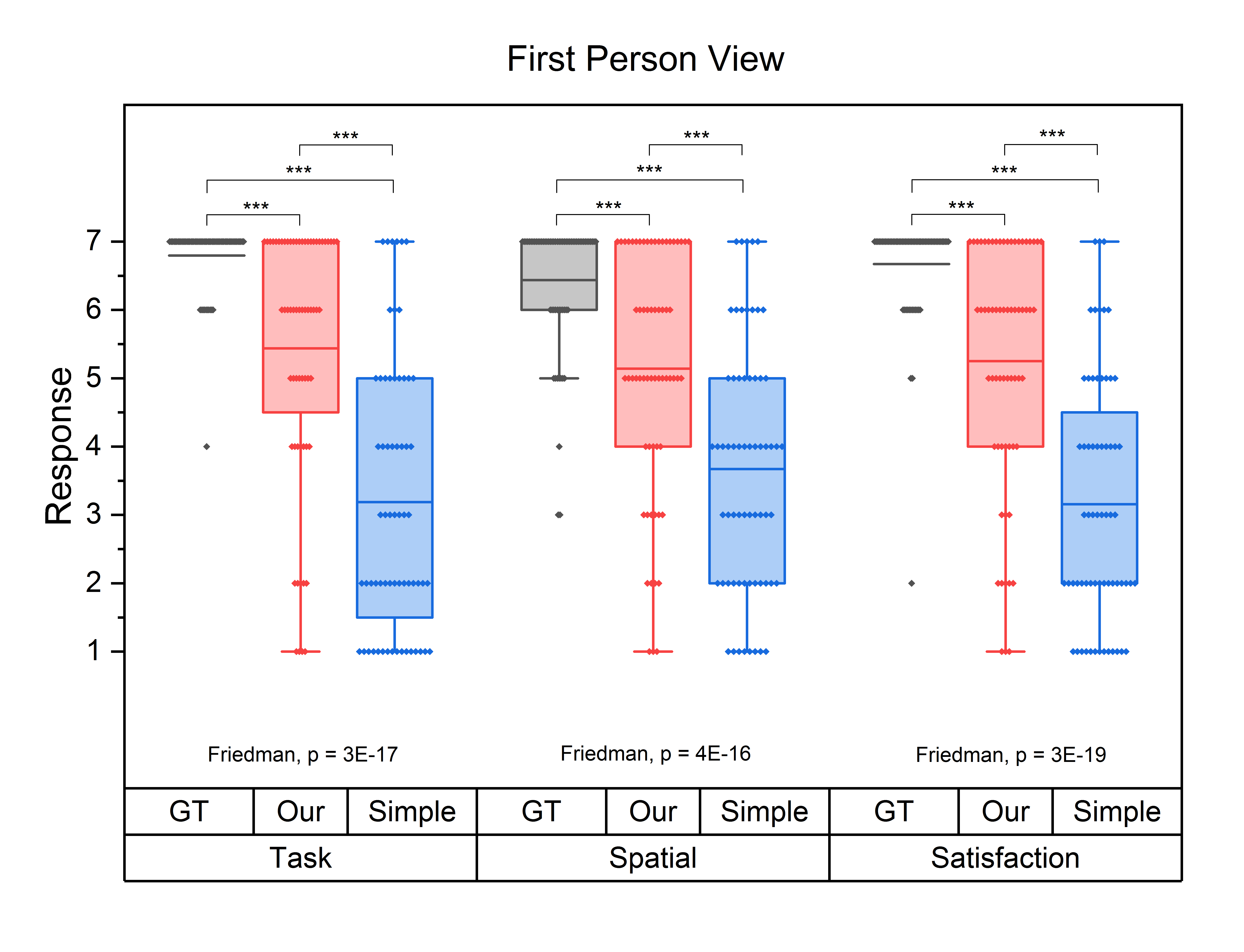}
 }
 \subfigure{
 \includegraphics[width=\columnwidth]{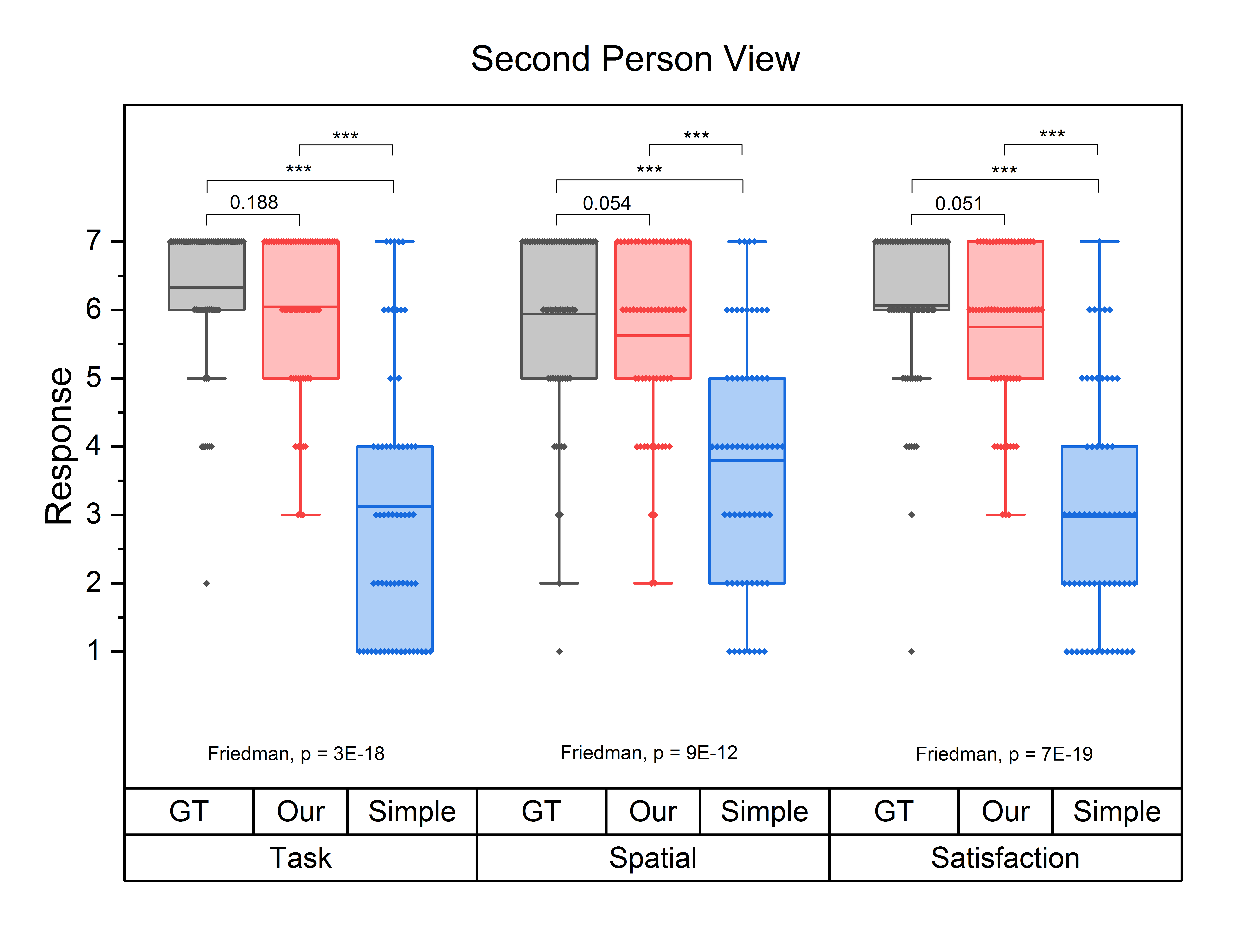}
 }
 \caption{Box plot results of first person view (top) and second person view (bottom) evaluation from EXP1.}
 \label{fig:sample}
\end{figure}

\subsection{User Study in AR Environment}
\label{sec:new_user_study}

We construct AR telepresence environments and conduct a two-part user study that consists of placement evaluation and system evaluation on presence and usability.
For the test, we set up two spaces with real furniture (TV, sofa, table, and chair) and modify the number and position of objects to make two different pair of spaces (a total of 4 configurations).
Figures \ref{fig:teaser}(b) and \ref{fig:AR_interaction} show snapshots of AR telepresence experience. 
In Figure \ref{fig:teaser}(b) both users are standing, and in Figure \ref{fig:AR_interaction} one is standing while the other is sitting. One can see that their avatars are placed appropriately to make interpersonal interaction while preserving the users' original poses. Supplementary video shows users experiencing AR telepresence in various positions within the test spaces.

The first user study (EXP1) evaluates the quality of placement of our method (\textbf{Our}), in comparison with a simple heuristic method (\textbf{Simple}) and a manual placement method (\textbf{GT}).
In the simple method, the avatar is placed simply to preserve the distance and angle between a user and an entity that is closest from the center of his/her visual field. If the corresponding avatar position is in collision with an object, its placement is modified to a nearby collision-free position found by random sampling.
In the manual placement method, the participant switches a view from his/her own AR space to the other party's space visualized with VR objects and move an avatar by hand-held controller to place avatar in ideal position and orientation. Refer to supplementary video for the detailed procedure.
The second user study (EXP2) evaluates the presence and usability of our automatic placement system \textbf{(Our : Automatic)} in comparison with the manual placement system \textbf{(GT : Manual)}. 
 
We recruited 14 female and 18 male students from local university community with an average age of 28.13 (SD=2.60). Their level of familiarity with VR or AR interfaces on a 7 point Likert scale from 1 to 7 was above the average (M=5.03, SD=1.40). All participants had an experience with VR or AR interfaces where 15 participants reported that they used a few times a month and 17 participants had used them a few times a year. Two participants conducted an experiment as a pair for both experiments. For EXP1, each participant from a pair acted as a main participant and a supporting participant in turn where the main participant was the evaluator. For EXP2, both participants evaluated the same questionnaire after using each system.

\subsubsection{EXP1: Placement Evaluation}
\label{sec:placement_evaluation}
The first experiment was a within-subject design where we investigated the effects of avatar placement given four tasks (Watch \textbf{TV, SIT} on chair, \textbf{LOOK} at avatar, \textbf{TALK} to avatar) on following criteria: task accomplishment \textbf{(TA)}, spatial similarity \textbf{(SS)} and overall satisfaction \textbf{(OS)}. The placement evaluation was two-fold involving the main participant and the supporting participant. First, the main participant evaluated the placement of self-avatar from the first person (egocentric) view. Second, the main participant evaluated the placement of the supporting participant's avatar from the second person view. 

\textbf{Procedure.} Both the main participant and the supporting participant performed each of four tasks in turn where the order of tasks was counter-balanced. For the first person view evaluation, after performing the given task in own AR space, the main participant switches the view of HMD to other party's VR space to place his/her own avatar and return back to AR space. Then the main participant knowing what the given task is evaluated the placement of avatar from the egocentric view of three different methods \textbf{(Our, Simple, GT)} presented in counter-balanced order. Three questions listed below are asked after the placement of each method. The participant rates each criteria on a 7-point Likert scale (1: Strongly Disagree - 7: Strongly Agree). For the second person view, the supporting participant places his/her own avatar in the same way while the main participant remains in VR space of the the supporting participant. After the avatar placement completed, the main participant returns back to his/her own AR space to evaluate the avatar placed by three different methods in counter-balanced order from the second person view. 
Questions on placement evaluation are as follows:
\begin{enumerate}
\item The avatar successfully accomplishes the given task (Task accomplishment).
\item The furniture configuration and arrangement around the avatar matches well with the my surrounding environment (Spatial similarity).
\item I am satisfied with the placement of the avatar given my placement (Satisfaction).
\end{enumerate}

\subsubsection{EXP2: System Evaluation}
\label{sec:system_evaluation}
The second experiment is also a within-subject design which we compare our automatic avatar placement system \textbf{(Our: Automatic)} with manual avatar placement system \textbf{(GT: Manual)} on presence and usability. 
The same size (16 participants) of new participants are recruited for the second experiment. The presence questionnaire is based on Networked Mind Measure of Social Presence \cite{harms2004internal} and Temple Presence Inventory \cite{lombard2009measuring}, while the questions for the system questionnaire are selected from USE Questionnaire \cite{lund2001measuring}.

\textbf{Procedure.} Two participants complete the same tasks in turn as in previous experiment. Instead of evaluating each placement as is done in EXP1, the participants observe the placement of other party's avatar and perform next given task. After using one of each system given in counter-balanced order, the participants are given two questionnaires listed in Tables \ref{tab:EXP2_presence_V} and \ref{tab:EXP2_usability_V} on presence and usability. The participants rate each question on a 7-point Likert scale (1: Strongly Disagree - 7: Strongly Agree). 

\begin{table*}[t]
\scriptsize
\begin{center}
\begin{tabular}{|c|c|c|c|c|c|c|c|c|c|c|c|c|c|}
\hline
\multirow{2}{*}{View} & \multirow{2}{*}{Method} & \multicolumn{3}{c}{TV} \vline & \multicolumn{3}{c}{SIT} \vline & \multicolumn{3}{c}{LOOK} \vline & \multicolumn{3}{c}{TALK} \vline \\
\cline{3-14}
 &  & T.A. & S.S & O.S. & T.A. & S.S & O.S. & T.A. & S.S & O.S. & T.A. & S.S & O.S.\\
\hline
\multirow{6}{*}{First} & \multirow{2}{*}{Our} & 4.69 & 4.56 & 4.5 & 5.25 & 4.81 & 5.13 & 5.88 & 5.81 & 5.75 & 6.06 & 5.62 & 5.88\\
& & (2.25) & (1.55) & (1.67) & (2.38) & (2.4) & (2.36) & (1.50) & (1.17) & (1.13) & (1.12) & (1.36) & (1.31)\\
\cline{2-14}
 & \multirow{2}{*}{GT} & 6.75 & 6.44 & 6.69 & 6.75 & 6.13 & 6.50 & 6.94 & 6.81 & 6.88 & 6.63 & 6.19 & 6.56\\
& & (0.45) & (0.81) & (0.48) & (0.77) & (1.26) & (0.73) & (0.25) & (0.40) & (0.34) &  (0.62) & (1.28) & (1.09)\\
\cline{2-14}
 & \multirow{2}{*}{Simple} & 2.25 & 2.69 & 2.25 & 1.94 & 3.25 & 2.50 & 5.06 & 4.44 & 4.31 & 3.31 & 4.38 & 3.44\\
& & (1.18) & (1.40) & (1.13) & (1.29) & (1.61) & (1.63) & (1.77) & (1.59) & (1.66) & (2.06) & (2.06) & (2.06)\\
\hline
\multirow{6}{*}{Second} & \multirow{2}{*}{Our} & 6.63 & 6.25 & 6.25 & 5.38 & 5.75 & 5.50 & 6.06 & 4.69 & 5.31 & 6.00 & 5.56 & 5.69\\
& & (0.80) & (0.93) & (1.06) & (1.54) & (1.44) & (1.27) & (1.06) & (1.66) & (1.49) & (1.32) & (1.26) & (1.14)\\
\cline{2-14}
 & \multirow{2}{*}{GT} & 6.75 & 6.44 & 6.50 & 6.19 & 6.31 & 6.19 & 5.89 & 5.19 & 5.31 &  6.63 & 6.00 & 6.31\\
& & (0.58) & (0.89) & (0.82) & (1.05) & (1.01) & (1.11) & (1.54) & (1.80) & (1.81) & (0.81) & (1.32) & (0.70)\\
\cline{2-14}
 & \multirow{2}{*}{Simple} & 2.94 & 3.69 & 2.69 & 2.50 & 3.69 & 2.63 & 4.13 & 4.13 & 3.75  & 3.13 & 3.63 & 2.94\\
& & (1.88) & (1.54) & (1.58) & (1.67) & (2.09) & (1.54) & (2.13) & (1.82) & (1.77) & (1.75) & (1.67) & (1.48)\\
\hline
\end{tabular}
\end{center}
\caption{Mean and standard deviation of rating on three criteria for each task from EXP1.}
\label{tab:EXP1_Stats}
\end{table*}

\begin{table*}[t]
\scriptsize
\begin{center}
\begin{tabular}{|c|c|c|c|c|c|c|c|c|c|c|c|c|c|}
\hline
\multirow{2}{*}{View} & \multirow{2}{*}{Value} & \multicolumn{3}{c}{TV} \vline & \multicolumn{3}{c}{SIT} \vline & \multicolumn{3}{c}{LOOK} \vline & \multicolumn{3}{c}{TALK} \vline \\
\cline{3-14}
 &  & T.A. & S.S & O.S. & T.A. & S.S & O.S. & T.A. & S.S & O.S. & T.A. & S.S & O.S.\\
\hline
\multirow{2}{*}{First} & Chi-square & 26.133 & 24.667 & 26.656 & 20.981 & 13.000 & 18.250 & 13.609 & 19.292 &	21.000 & 17.815 & 12.378 & 22.627\\
\cline{2-14}
 & p & *** & *** & *** & *** & ** & *** & ** & *** & *** & *** & *** & ***\\
\hline
\multirow{2}{*}{Second} & Chi-square & 25.064 & 21.708 & 21.709 & 26.772 & 18.681 & 26.772 & 10.292 & 4.204 & 8.739 & 20.039 & 13.440 & 26.920\\
\cline{2-14}
 & p & *** & *** & *** & *** & *** & *** & ** & 0.122 & * & *** & ** & *** \\
\hline
\end{tabular}
\end{center}
\caption{Friedman test results for each task from EXP1. (* $<$ 0.05, ** $<$ 0.01, *** $<$ 0.001)}
\label{tab:EXP1_Friedman}
\end{table*}

\begin{table*}[t]
\scriptsize
\begin{center}
\begin{tabular}{|c|c|c|c|c|c|c|c|c|c|c|c|c|c|}
\hline
\multirow{2}{*}{View} & \multirow{2}{*}{Comparison} & \multicolumn{3}{c}{TV} \vline & \multicolumn{3}{c}{SIT} \vline & \multicolumn{3}{c}{LOOK} \vline & \multicolumn{3}{c}{TALK} \vline \\
\cline{3-14}
 &  & T.A. & S.S & O.S. & T.A. & S.S & O.S. & T.A. & S.S & O.S. & T.A. & S.S & O.S.\\
\hline
\multirow{3}{*}{First} & Our vs GT & ** & * & ** & 0.148 & 0.173 & 0.182 & * & * & ** & 0.101 & 0.364 & 0.143\\
\cline{2-14}
 & Our vs Simple & * & * & * & ** & 0.132 & * & 0.327 & * & * & ** & * & **\\
\cline{2-14}
 & GT vs Simple & *** & *** & *** & ** & ** & ** & ** & ** & ** & ** & * & **\\
\hline
\multirow{3}{*}{Second} & Our vs GT & 1.000	& 0.771	& 1.000	& 0.255	& 0.102	& 0.137	& 1.000 & 0.839 & 1.000 & 0.142 & 0.173	& 0.069\\
\cline{2-14}
 & Our vs Simple & ** & ** & ** & ** & ** & ** & * & 0.837 & * & ** & 0.066 & **\\
\cline{2-14}
 & GT vs Simple & ** & ** & ** & *** & ** & *** & 0.147	& 0.126 & *	& ** & * & **\\
\hline
\end{tabular}
\end{center}
\caption{Post-hoc analysis for each task from EXP1: Wilcoxon Signed Rank Test (Pair-wise) with Bonferroni correction (* $<$ 0.05, ** $<$ 0.01, *** $<$ 0.001).}
\label{tab:EXP1_Wilcoxon}
\end{table*}

\subsubsection{Experiment Results and Analysis}
\label{sec:experiment_results}
\subsubsection*{EXP1: Placement Evaluation}
We performed non-paramateric Friedman tests for all measures between three methods based on the criteria for each task for two view points. For post-hoc analysis, we used a pairwise Wilcoxon signed-rank test with Bonferroni correction. Table \ref{tab:EXP1_Stats} shows the mean and standard deviation on responses for all measures.

Overall, Our method is significantly better than Simple method, but significantly worse than GT from first person view, and it is competitive with GT and significantly better than Simple method from second person view (Figure \ref{fig:sample}). Our interpretation is that the participants were sensitive about the difference from GT acquired by him/herself in first person view, but for second person view, which is actually the case of practical application of our method, the participants were satisfied with our method in terms of three criteria.

Table \ref{tab:EXP1_Friedman} shows that we found significant differences on all measures except for spatial similarity on LOOK task from second person view. 
Post-hoc analysis is provided in Table \ref{tab:EXP1_Wilcoxon}. 
For TV task, Our method was not significantly different from GT from second person view while GT scored significantly higher for first person view for all three criteria. For viewing tasks, viewing angle seems to be considered important as it tends to be preserved well in GT while less so in Our method. However, for the second person view, both GT and Our method were able to produce high scores because the participants were not sensitive to other party's viewing angle.

For SIT task, post-hoc tests were not significant for GT and Our method from both first and second person view. For both views, as long as the avatar was placed in the chair or sofa, the scores were high for both GT and Our method. For the spatial similarity from first person view, Simple method was not significantly different from Our method as it was often able to place the avatar near a chair with a table whichever was close to the center of gaze of the participant.

For LOOK task from first person view, post-hoc tests showed that GT was significantly better than Our method at looking at other party's avatar, and Simple method was not significantly different with Our method on task accomplishment. In terms of second person view, Our method was not significantly different from GT on all criteria and Simple method was not significantly different from GT on task accomplishment and spatial similarity. One possible interpretation is that all methods were able to place avatar's orientation toward the main participant so that the scores for Simple method were somewhat comparable to other methods.

For TALK task, Our method was not significantly different from GT. Simple method was not significantly different from Our method for spatial similarity, but its task accomplishment and satisfaction scores were significantly lower than other two methods. Our interpretation is that the participants were quite sensitive about small difference in distance and orientation in Talk task.

\subsubsection*{EXP2: System Evaluation}

We performed pairwise Wilcoxon signed-rank test for each question from questionnaires. For presence questionnaire, we found significant difference in favor of Manual system (GT) with Questions 1 and 2  while differences were not significant for the remaining questions (Table \ref{tab:EXP2_presence_V}). Our interpretation is that the participants were able to visit other party's VR space during the manual placement and understand actual context of the avatar so that the participants can focus more on partner during the interaction. However, the participants were able to understand other party under the given task with good presence using our automatic system.

\begin{table}[t]
\scriptsize
\begin{center}
\begin{tabular}{|l|c|c|c|}
\hline
Question & Our & GT & P\\
\hline
1. I remained focused on my partner  & \multirow{2}{*}{5.00 (1.10)}  & \multirow{2}{*}{5.56 (1.03)} & \multirow{2}{*}{*}\\
\hspace{0.25cm}throughout our interaction. &  &  & \\
\hline
2. My partner remained focused on & \multirow{2}{*}{4.69 (1.01)} & \multirow{2}{*}{5.38 (0.96)} & \multirow{2}{*}{*}\\
\hspace{0.25cm}me throughout our interaction. & & & \\
\hline
3.  My partner's thoughts were clear.  & \multirow{2}{*}{5.31 (1.49)} & \multirow{2}{*}{5.00 (0.89)} & \multirow{2}{*}{0.289}\\
\hspace{0.25cm}to me. & & & \\
\hline
4.  It was easy to understand my  & \multirow{2}{*}{5.13 (1.15)} & \multirow{2}{*}{5.06 (1.24)} & \multirow{2}{*}{0.864}\\
\hspace{0.25cm}partner. & & & \\
\hline
5.  It seems that my partner had come  & \multirow{2}{*}{5.50 (1.10)} & \multirow{2}{*}{5.56 (1.21)} & \multirow{2}{*}{0.963}\\
\hspace{0.25cm}to the place I was. & & & \\
\hline
6.  It seems that I and my partner was  & \multirow{2}{*}{5.38 (0.96)} & \multirow{2}{*}{5.50 (1.26)} & \multirow{2}{*}{0.476}\\
\hspace{0.25cm}together in the same place. & & & \\
\hline
\end{tabular}
\end{center}
\caption{Presence questionnaire results - Mean and standard deviation for each condition and p-value in EXP2 (* $<$ 0.05).}
\label{tab:EXP2_presence_V}
\end{table}

For usability, we found significant difference in favor of automatic system (Our) with Questions 5, 6 and 11 (Table \ref{tab:EXP2_usability_V}). This result indicates that Our automatic system is considered easier and simpler to use, and the participants want to recommend Our system to a friend more than the manual system.

\begin{table}[t]
\scriptsize
\begin{center}
\begin{tabular}{|l|c|c|c|}
\hline
Question & Our & GT  & P\\
\hline
1. It helps me be more effective. & 5.18 (1.33) & 4.56 (1.37) & 0.160\\
\hline
2. It is useful. & 5.19 (1.11) & 4.88 (1.26) & 0.751\\
\hline
3.  It meets my needs. & 4.94 (1.48) & 4.63 (1.71) & 0.755\\
\hline
4.  It is easy to use. & 5.63 (1.71) & 4.69 (1.40) & 0.064\\
\hline
5.  It is simple to use. & 6.06 (1.12) & 4.94 (1.53) & **\\
\hline
6.  It is user friendly. & 5.31 (1.58) & 4.31 (1.49) & *\\
\hline
7.  I learned to use it quickly. & 5.81 (1.33) & 5.44 (1.03) & 0.141\\
\hline
8.  I easily remember how to use it. & 6.00 (1.55) & 5.25 (1.00) & 0.050\\
\hline
9.  It is easy to learn to use it. & 5.44 (1.55) & 5.25 (1.00) & 0.524\\
\hline
10.  I am satisfied with it. & 4.63 (1.31) & 4.44 (1.36) & 0.642\\
\hline
11.  I would recommend it to a friend. & 4.81 (1.38) & 3.94 (1.18) & *\\
\hline
12.  It is fun to use. & 4.81 (1.72) & 4.81 (1.56) & 1.000\\
\hline
13.  It works the way I want it to work. & 4.75 (0.86) & 5.13 (1.63) & 0.175\\
\hline
\end{tabular}
\end{center}
\caption{Usability questionnaire results - Mean and standard deviation for each condition and p-value in EXP2 (* $<$ 0.05, ** $<$ 0.01).}
\label{tab:EXP2_usability_V}
\end{table}

\subsubsection{Observation and General Feedback}
\label{sec:experiment_feedback}
For the placement evaluation (EXP1), participants found that our method works well for the context from the given tasks especially from the second person view. Some participants noted that acquiring GT affects memory on the first person view, which has a favorable effect on GT method in the first person view. However, from the second person view where the GT placement was acquired by the supporting participant, the evaluation from the main participant was not significantly different between GT and Our method. 

For the system evaluation (EXP2), many participants mentioned that Our system was easy and fun to use and they were surprised that the automatic placement was actually supporting the given tasks very well. On the other hand, some participants also noted that GT system that allows for exploring other party's space was also interesting. Many participants said that manual placement with input device was time-consuming while some participants enjoyed the process.

Overall, most participants expressed that they had a hard time from a heavy device with VR sick as the experiment went on for a while. One participant suggested that voice chatting would be helpful for real application, but our focus was mainly on evaluating placement this time.

\section{Discussion and Future Work}
\label{sec:discussion}
In this paper, we presented a method for placing avatars to preserve the meaning of various aspects of user's placement in a dissimilar indoor environment. A core technical component for this was the neural network that estimates the similarity between two placements from different spaces, trained based on people's preference for avatar positions obtained through a user survey. We have shown the effectiveness of our method through experiments with a VR and AR-based telepresence system. In this section, we discuss several limitations of our work and important future research themes.

First, we used only low-level features, such as geometry and object category information, as inputs to our system. With this approach, however, it was difficult for the avatar to preserve the meaning of the location intended by the user when it can be interpreted in various ways (i.e., when the avatar of the other party and many objects are in the vicinity). Recognizing user's actual attention will greatly help avatar generate motions that reflect the user's intended movement. 
{\blue It would be also worth exploring alternative feature models than ours (e.g., more sophisticated spatial features that can distinguish subtle differences of furniture arrangement) that can improve the placement quality. }

Next, we conducted a user survey on a rather limited number of space pairs. As shown in \ref{fig:UserStudyNNDistance}, the data shows a statistically significant distribution, and we were able to apply the trained model to indoor spaces from other datasets. Nevertheless, more extensive user survey data on a wide variety of spaces and human-avatar placements will help achieve a more generalized avatar placement method,
{\blue for example by training neural networks in an end-to-end way to discover placement-related features rather than manually specifying features as done in this work.}

In order to realize the avatar-based telepresence to be used in the real world, additional progress in many directions needs to be made with regard to the avatar animation. 
In particular, generating avatar's locomotion as well as gesture and gaze animation to preserve the remote user's movement semantics remains as an important and interesting future research direction.

Lastly, in this work we assumed two users, among whom only one moves at a time. Further research is necessary to accommodate more than two concurrently moving users.

\ifCLASSOPTIONcompsoc
  \section*{Acknowledgments}
\else
  \section*{Acknowledgment}
\fi
This work was partly supported by Samsung Science and Technology Foundation (SRFC-IT1701-14).

\ifCLASSOPTIONcaptionsoff
  \newpage
\fi

\bibliographystyle{IEEEtran}
\bibliography{refrences.bib}

\begin{IEEEbiography}[{\includegraphics[width=1in,height=1.25in,clip,keepaspectratio]{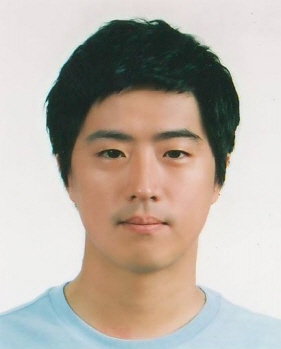}}]{Leonard Yoon} is a Ph.D. candidate with the Graduate School of Culture Technology at KAIST. He received the M.S. degree in Electrical Engineering from KAIST, Korea, in 2014 and B.S. degree in Electrical and Computer Engineering from University of California, San Diego, USA, in 2009. His research interests include 3D scene understanding and AR telepresence.
\end{IEEEbiography}
\vskip -2\baselineskip plus -1fil
\begin{IEEEbiography}[{\includegraphics[width=1in,height=1.25in,clip,keepaspectratio]{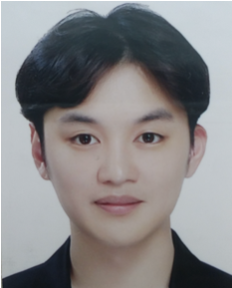}}]{Dongseok Yang} is a Ph.D. candidate with the Graduate School of Culture Technology at KAIST. He received the M.S. degree in Culture Technology from KAIST, Korea, in 2020 and B.S. degree in Multimedia Engineering from Dongguk University, Korea, in 2018. His research interests include real-time character animation synthesis and AR telepresence.
\end{IEEEbiography}
\vskip -2\baselineskip plus -1fil
\begin{IEEEbiography}[{\includegraphics[width=1in,height=1.25in,clip,keepaspectratio]{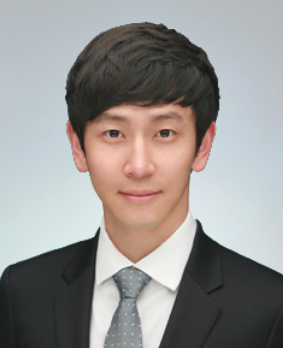}}]{Jaehyun Kim} is a Ph.D. candidate with Program of Brain and Cognitive Engineering in Department of Bio and Brain Engineering at KAIST. His research interests include brain-inspired artificial intelligence for spatial recognition, path finding, and sequential decision making. He received the M.S. degree in Culture Technology from KAIST, in 2016 and B.S. degree in Media / Information and Computer Engineering from Ajou University, in 2011.
\end{IEEEbiography}
\vskip -2\baselineskip plus -1fil
\begin{IEEEbiography}[{\includegraphics[width=1in,height=1.25in,clip,keepaspectratio]{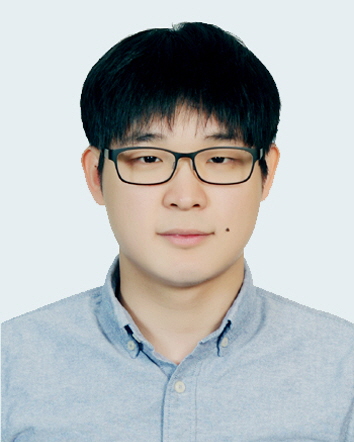}}]{Choongho Chung} is a Ph.D. candidate with the Graduate School of Culture Technology at KAIST. He received the M.S degree in Culture Technology from KAIST, Korea, in 2018 and a B.S degree in Mechanical Engineering in KAIST, Korea, in 2014. He is interested in telepresence applications, human attention recognition, and generation of character animation.
\end{IEEEbiography}
\vskip -2\baselineskip plus -1fil
\begin{IEEEbiography}[{\includegraphics[width=1in,height=1.25in,clip,keepaspectratio]{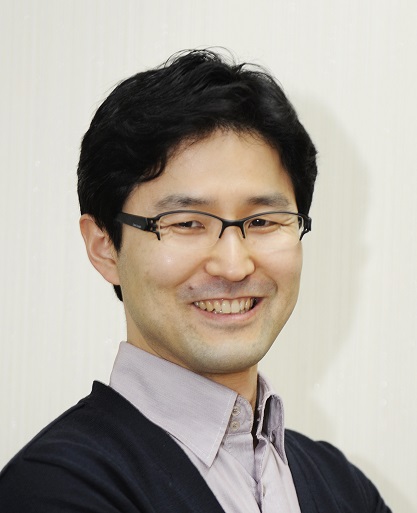}}]{Sung-Hee Lee} is an Associate Professor with the Graduate School of Culture Technology at KAIST. His research interests include autonomous human animation, avatar motion generation, and human modeling. He received the Ph.D. degree in Computer Science from University of California, Los Angeles, USA, in 2008, and the B.S. and the M.S. degree in Mechanical Engineering from Seoul National University, Korea, in 1996 and 2000, respectively.
\end{IEEEbiography}

\end{document}